\documentclass[prd,a4paper,eqsecnum,nofootinbib,preprint,floatfix]{revtex4}
\usepackage{colordvi}
\usepackage{graphicx}
\usepackage{bm}
\usepackage{multirow}
\usepackage{enumerate}
\reversemarginpar
\def\<{\langle}
\def\>{\rangle}
\newcommand{\text}{\rm}
\def\Tr{{\rm Tr}\,}

\def\Eq#1{Eq.~(\ref{#1})}

\renewcommand\Re{{\text Re}\,}

\def\tilde{\widetilde}

\def\tilde{\widetilde}
\def\tQ{{\tilde Q}}
\def\tP{{\tilde P}}

\begin{document}

\begin{titlepage}

\vspace*{0.7in}
 
\begin{center}
{\large\bf Applying the Wang-Landau Algorithm to
 Lattice Gauge Theory\\}
\vspace*{1.0in}
{Barak Bringoltz and Stephen R. Sharpe\\
\vspace*{.2in}
Department of Physics, University of Washington, Seattle,
WA 98195-1560, USA\\
}
\end{center}

\vspace*{0.55in}

\begin{center}
{\bf Abstract}
\end{center}

We implement the Wang-Landau algorithm in the
context of $SU(N)$ lattice gauge theories. We study the
quenched, reduced version of the lattice theory and calculate its
density of states for $N=20,30,40,50$. 
 We introduce a variant of the original algorithm in which the weight
function used in the update does not asymptote to a fixed function, 
but rather continues to have small fluctuations which enhance tunneling.
We formulate a method to evaluate
the errors in the density of states, and use the result to calculate
the dependence of the average action density and the specific heat on
the `t Hooft coupling $\lambda$. This allows us to locate the coupling
$\lambda_t$ at which a strongly first order transition occurs in the
system. For $N=20$ and $30$ we compare our results to those obtained
using Ferrenberg-Swendsen multi-histogram reweighting and find
agreement with errors of $0.2\%$ or less. Extrapolating
our results to $N=\infty$ we find $\left(\lambda_t\right)^{-1} =
0.3148(2)$. We remark on the significance of this result for the
validity of quenched large-$N$ reduction of $SU(N)$ lattice gauge
theories.

\end{titlepage}

\setcounter{page}{1}
\newpage
\pagestyle{plain}

\section{Introduction}
\label{section_intro}

Strongly first order phase transitions provide a difficult
challenge for numerical simulations. 
Consider, for example, a physical system whose
interactions are characterized by a single coupling $g$. A naive
approach to estimate the transition coupling, $g_t$, is to perform
Monte-Carlo (MC) simulations at couplings that are close to $g_t$ and
locate the point at which different observables are
discontinuous. These measurements, however, are affected by a
strong hysteresis whose width grows with the number of degrees of freedom
$N_{\rm dof}$. For large enough $N_{\rm dof}$, this width
dominates the error in the transition coupling,
which can result in large uncertainties 
($10\%-20\%$ in the example we consider here). 

To obtain improved precision the way forward is undoubtedly to
use reweighting algorithms.
For example, Ferrenberg-Swendsen reweighting (FSR) 
uses MC simulations to measure the
normalized histogram of the action $A$
for a coupling $g=g_0$ which is close to $g_t$~\cite{FSR}. 
By construction, this is given by
\begin{equation}
h_0(A) \sim \rho(A) \times P_{\rm Boltzmann}(g_0;A)\,, 
\label{h_0}
\end{equation}
where $\rho(A)$ is the density of states and 
$P_{\rm Boltzmann}(g_0;A)$ is the Boltzmann weight.
In the $SU(N)$ lattice gauge theories we consider here, 
$g$ is typically identified with the bare lattice 't Hooft 
coupling $\lambda$, and 
$P_{\rm Boltzmann}(g_0;A) \sim \exp (A/\lambda)$.
A measurement of $h_0(A)$ thus provides
an estimate for $\rho(A)$. Using this, one can
estimate the histogram $h_{\lambda}(A)$ at any other coupling
by reweighting:
\begin{equation}
h_{\lambda}(A) \sim h_0(A) \times 
\exp \left(A/\lambda - A/\lambda_0\right)\,.
\end{equation}
One then determines 
the coupling $\lambda_t$ at which the corresponding histogram
$h_{\lambda_t}(A)$ takes a double-peak form, as expected for a
first-order phase transition. 
In practice this amounts to calculating the average action ${\cal A}$ and its associated
specific heat ${\cal C}$ as a function of $\lambda$
\begin{eqnarray}
{\cal A}(\lambda) &=& \int \,\, dA \,\, h_{\lambda}(A) \,\, A, \label{av_action}\\
{\cal C}(\lambda) &=& \int \,\, dA \,\,h_{\lambda}(A) \,\, 
(A-{\cal A}(\lambda))^2,
\label{spec_heat}
\end{eqnarray}
and finding the coupling $\lambda_t$ at which ${\cal C}(\lambda)$
peaks.

The FSR method has an obvious shortcoming.
When reweighting from $\lambda_0$ to $\lambda$ one is ``amplifying'' the
contribution to $\rho(A)$ from field configurations that are important
at $\lambda$ while suppressing those relevant at $\lambda_0$.  
If, however, the field configurations probed at
$\lambda_0$ are substantially different from those important at
$\lambda$, then this amplification can be dominated by statistical
noise. This ``overlap'' problem can cause a large systematic error which
may be hard to evaluate. To avoid it one needs to ensure that the
field configurations that are important at $\lambda$ are reasonably
sampled when performing measurements at $\lambda_0$. In ordinary
situations this means that the couplings $\lambda_0$ and $\lambda$
need to be sufficiently close.  When $\lambda\simeq \lambda_t$,
however, there are field configurations which are very hard to
probe. These are the tunneling configurations between the two
phases. Thus for reweighting to work in the context of locating a
strongly first order phase transition, one requires that a sufficient
number of tunneling events are observed whilst measuring the
histograms.  This requirement
can be very restrictive when $N_{\rm dof}$ is large because the
tunneling probability typically falls exponentially
as $N_{\rm dof}$ increases.
Consequently, when performing reweighting, it
is crucial to use an algorithm that encourages tunneling events.

In this paper we do not discuss all the different alternatives to FSR
(which are discussed, for example, in Ref.~\cite{Berg-Book}).\footnote{%
%
One attractive option is the multi-canonical algorithm
of Ref.~\cite{MUCA}. We did not use this approach
because it had been found in Ref.~\cite{Campostrini},
which studied a model similar to ours,
that a very delicate tuning of
parameters was needed for large enough systems. 
One advantage of the WL algorithm is that it is self-tuning.
An alternative, applied successfully in Ref.~\cite{BB}, is to
use the WL algorithm to provide an estimate of the weight function
of the multi-canonical algorithm.}
%
Instead we choose to study (a variant of) the Wang-Landau (WL)
reweighting algorithm, which was introduced in the field of
statistical mechanics~\cite{WLR}, and is particularly well
suited for promoting tunneling. In the context of gauge theories,
this algorithm can be considered to be a modern incarnation of the
early attempts, such as the ones in Ref.~\cite{Bhanot_DOS}, to
calculate $\rho(A)$ of lattice gauge theories 
(see also Refs.~\cite{FKS,Denbleyker-et-al}). 


A sketch of the WL algorithm is as follows (a more precise definition
will be given in Sec.~\ref{sec_WLR}). From here on we use the action
{\em density} $E\sim A/N_{\rm dof}$ as our prime observable, and so
denote the density of states by $\rho(E)$. We denote the Monte-Carlo
time by $t$ and the WL estimate of the density of states at time $t$ by
$\rho_t(E)$. 

\begin{enumerate}
\item Begin at MC-time $t=0$ with
an initial estimate for the density of states, $\rho_{0}(E)$.
\item Use $1/\rho_{t}(E)$ as a Boltzmann weight to create a series of field configurations.
\item Update $\rho_{t}(E)\to \rho_{t+1}(E) = \rho_{t}(E) + \delta
\rho(E)$. The update function $\delta \rho(E)$ depends on the MC
history between times $t$ and $t+1$ in a way that biases {\em against} small
or null changes of $E$ at time $t+1$, and so encourages tunnelings. 
This is an essential point in Wang-Landau
reweighting (WLR) and we
discuss it in greater detail in Section~\ref{sec_WLR}.
\item GoTo step (2).
\end{enumerate}

One can show, with some assumptions,
that for large enough $t$,
$\rho_{t}(E)$ converges to the vicinity of $\rho(E)$, 
and subsequently fluctuates around it.
We provide this demonstration in Sec.~\ref{sec_WLA_theory},
generalizing the discussion in Ref.~\cite{ZB}.
The fluctuations are an intrinsic part of the WL algorithm,
and are the consequence of the ergodicity enforced
by the ``biasing'' in step (3) above.
Once converged, the algorithm generates
a chain of field configurations that
are weighted by an approximately flat probability function
\begin{equation}
P(E) \sim \rho(E) \times 1/\rho_{t}(E) \approx E-{\rm
independent}\,. \label{P_flat}
\end{equation}
Consequently, all values of $E$ will be accessed with approximately equal
probability, including those corresponding to tunneling
events.
%
%
Using the estimate of $\rho(E)$ one can calculate
the specific heat ${\cal C}(\lambda)$ and locate its peak.

In this paper we adapt the WL algorithm to $SU(N)$ gauge theories and
in particular formulate a systematic way to evaluate errors
 in derived quantities such as ${\cal C}(\lambda)$. 
The model we choose to study is obtained from four-dimensional $SU(N)$
lattice gauge theories by ``quenched reduction'' to a single
lattice site (see, for example, Refs.~\cite{QEK_old_papers} 
and the recent review in Ref.~\cite{QEK_paper}). 
It is a matrix model of four $SU(N)$
matrices. The interactions between these matrices are governed by the
`t Hooft coupling $\lambda$, and lead to a nontrivial change in
various expectation values as one moves from
strong to weak couplings. This behavior becomes a strongly first order
transition when $N\to\infty$ and it is this transition we wish to
analyze using the WL algorithm.

As noted above, we use a variant of the WL algorithm.
The key difference between our variant and
the original WL algorithm (``WL$^0$''),
is that the latter
includes an iterative procedure which we do not use.\footnote{%
A less important difference is that one must adapt the original algorithm
from systems with discrete variables to those with continuous degrees of
freedom. We describe how this has been done below.
}
Namely, in WL$^0$, the steps (1-4) above are first applied with a
given update function, $\delta\rho_1(E)$, for some Monte-Carlo time
$T_1$.  The time $T_1$ is determined ``on the fly'' by requiring that
the values of $E$ that are visited are sufficiently uniform.
Once the chosen criterion is fulfilled, the function 
$\delta \rho_1(E)$ is replaced by $\delta \rho_2(E)$ which is smaller,
i.e. obeys $|\delta \rho_2(E)|< |\delta \rho_2(E)|$.
The procedure is then iterated until
the magnitude of $\delta \rho(E)$ drop below the machine accuracy. As
shown in \cite{Trebst1}, and discussed below (for example see
Section~\ref{sec_tune}), the tunneling rate, which is what one wishes
to increase in the WL algorithm, decreases as
the size of $\delta \rho(E)$ is decreased, 
making the WL$^0$ less and less efficient as
it is iterated. For this reason we keep $\delta \rho(E)$ finite,
and thus always have a Boltzman weight which varies (albeit by
a small amount) so as to maintain the tunneling rate. This also avoids
the need to tune extra parameters, such as the choice of the flatness
criterion. Despite the lack of a fixed weight, we can measure expectation
values since $\rho_t(E)$ fluctuates around the correct value.

A different solution to the tunneling problem, involving ultimately fixed weights,
is presented in Ref.~\cite{Trebst2}.


The outline of the paper is as follows. We first introduce the matrix
model that we study in Sec.~\ref{sec_QEK}. We describe the
Wang-Landau algorithm and its properties in Sec.~\ref{sec_WLR}, and in
Sec.~\ref{sec_tune} we describe our implementation
and in particular the tuning of parameters and
the calculation of errors.
In Sec.~\ref{sec_res} we report our
results and compare them to corresponding data obtained using
Ferrenberg-Swendsen reweighting and standard Monte-Carlo
simulations. We summarize in Sec.~\ref{sec_summary}, and
remark on the implication of our results to the validity of large-$N$
quenched reduction of $SU(N)$ lattice gauge
theories. Appendix~\ref{app_alg} includes a description of the
different update algorithms which we use, and Appendix~\ref{app_more}
discusses additional technical issues related to the implementation of the
Wang-Landau algorithm.

Our results for the transition coupling
$\lambda_t$ were already quoted in Ref.~\cite{QEK_paper}.

\section{Quenched-reduced $SU(N)$ lattice gauge theories}
\label{sec_QEK}

In this section we briefly describe the matrix model which we study.
For a discussion of its relevance to $SU(N)$ gauge theories
we refer to Ref.~\cite{QEK_paper} and references therein.

\subsection{Definition of the matrix model}
\label{model_def}

The model consists of four $SU(N)$ matrices $\left\{V_{\mu} \,
; \, \mu=1,2,3,4\right\}$. Observables are built from the 
$SU(N)$ `link matrices' $U_\mu$ defined by
\begin{equation}
U_\mu \equiv V_\mu \Lambda_\mu V^\dag_\mu,
\end{equation}
where $\Lambda_\mu$ are the fixed, diagonal $SU(N)$ matrices
\begin{equation}
\Lambda^{ab} \equiv \delta^{ab} \, e^{ip^a_\mu} \,:
\qquad p^a_\mu \in [0,2\pi]\,,\qquad a,b\in [1,N]\,.
\end{equation}
The quenched momenta $p^a_\mu$ are drawn from some distribution---various
possibilities are discussed in Ref.~\cite{QEK_paper}.
Since our focus here is on the algorithm, we pick one choice
of momenta (the ``clock'' momenta),
\begin{equation}
p^a_\mu = \frac{2\pi}{N} \left( a - \frac{N+1}2\right) \quad ;\quad a
\in [1,N]\,,
\label{clock}
\end{equation}
and use it throughout.
Expectation values of an observable ${\cal O(U)}$ are calculated 
via
\begin{equation}
\left\langle{{\cal O(U)}}\right\rangle \equiv Z(b)^{-1} \int \prod_\mu
DV_\mu \,\,\exp{\left(\,b\,A\right)} \,\, {\cal O(U)} \,,
\label{eq:O(U)theta}
\end{equation}
where here the action $A$ is
\begin{equation}
A = N \sum_{\mu<\nu} 2 {\rm Re}\,\Tr\left(U_\mu U_\nu U_\mu^\dagger
U_\nu^\dagger \right) \,,
\label{eq:SQEK}
\end{equation}
and $b$ is the inverse of the `t Hooft coupling, $b=1/\lambda$.  The
partition function $Z(b)$ is
\begin{equation}
Z(b) \equiv \int \prod_\mu DV_\mu \,\exp(\,b\,A)\,,
\label{eq:ZQEK}
\end{equation}
and $DV_\mu$ is the Haar measure on $SU(N)$. The integral
over $V_\mu$ includes matrices that realize permutations in the
indices $a$ of $p^a_\mu$, and so the construction above is invariant
under such permutations. Thus one can equally define the model with any
set of $p^a_\mu$ obtained from \Eq{clock} by permuting the $a$
indices, independently in each direction.

We take the action density to be
\begin{equation}
E \equiv \frac{A}{12N^2}\,,
\end{equation}
so that it is the average, normalized plaquette.
We consider, for simplicity,
only even values of $N$, for which $E$ lies in the range $[-1,1]$.

\subsection{A sketch of the phase diagram}
\label{sketch_PD}

\begin{figure}[htb]
\includegraphics[width=10cm,angle=-90]{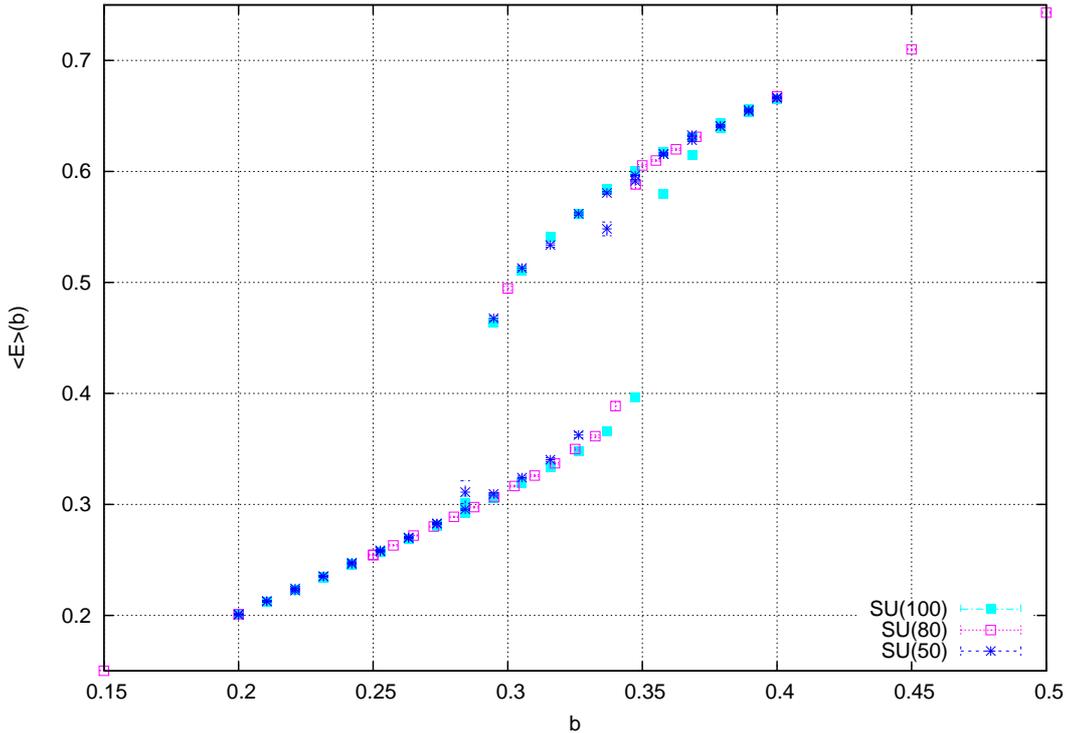}
\caption{Hysteresis plots of the average action density $E$
versus $b=1/\lambda$ 
for $SU(50)$ ([blue] crosses), $SU(80)$ ([magenta] open
squares), and $SU(100)$ ([light blue] filled squares). For details 
see Ref.~\cite{QEK_paper}.
\label{hyst_map}}
\end{figure}

In Ref.~\cite{QEK_paper} we mapped the phase diagram of the model in $b$,
and saw strong evidence that there exists a first order phase
transition at $b=b_t \simeq 0.3$. This was also seen in earlier
studies of the model (for example in Ref.~\cite{Okawa_QEK}). To
demonstrate this we present in Fig.~\ref{hyst_map} our results for
$\langle E\rangle (b)$, obtained using conventional MC simulations
(using algorithms described in Appendix.~\ref{app_more}).
A clear hysteresis is seen, with width increasing with $N$,
as expected since $N_{\rm dof}\propto N^2$.
Our aim in this paper is to develop a method that can accurately
 locate the coupling $b_t$ at which this transition occurs.

\section{Wang-Landau reweighting}
\label{sec_WLR}

Reweighting methods start by integrating out all
but a few variables (usually one or two) from the partition function. 
We use a single remaining variable, the action density $E$:
\begin{eqnarray}
Z(b) &=& \int DV \, \exp{\left(b\,A\right)} \equiv \int_{-1}^1 dE \,
\rho(E) \exp{\left(12 N^2 \,b \, E\right)} \\ &\equiv& \int dE \exp
\left(\omega(E) + 12 N^2 \, b\, E \right).
\end{eqnarray}
Here $\rho(E)$ is the number density of field configurations
with action density in the range $[E,E+dE]$ and $\omega(E)$ is the
associated ``entropy'':
\begin{equation}
\omega(E) = \log \, \left( \rho(E)\right).
\end{equation}
The expectation value of an observable, ${\cal O}(E)$, 
that depends solely on $E$ can be written as
\begin{equation}
\<{\cal O}\> = Z^{-1}(b)\int dE \, \exp \left(\omega(E) + 12 \, N^2\,
b\, E \right)\, {\cal O}(E). \label{OE}
\end{equation}
Thus, calculating the function $\omega(E)$ can be considered as
solving the theory in the sector that couples to operators of the form
${\cal O}(E)$.

The original WL algorithm was introduced in Ref.~\cite{WLR} 
to study statistical systems with discrete degrees of freedom. 
$E$ then takes discrete values, and this is reflected in the formulation
of the original algorithm. In our case, however, $E$ is continuous, and we 
need to adapt WLR accordingly. Two alternative approaches have
been considered in the statistical-mechanics and molecular-dynamics
literature:
(1) Discretize $E$ into bins and then apply the
discrete WL algorithm. This approach has been used, for example, 
to study the classical Heisenberg model~\cite{WLD}; 
(2) Generalize the WL algorithm 
so that it updates $\omega(E)$ treating $E$ as a 
continuous variable~\cite{pre_ZSTL,ZSTL}.\footnote{%
Binning is still required to store functions like $\omega(E)$ in memory, but
does not play an essential role in the algorithm.
See Appendix~\ref{app_more} for further discussion.}
Based on preliminary studies, 
we chose to pursue only option (2) in detail.
We follow and extend the approach suggested in Ref.~\cite{ZSTL},
which we next describe and analyze in some detail.

\subsection{The algorithm}
\label{sec_WLA}

The algorithm proceeds by updating an estimate of the entropy,
$\omega_t(E)$, where $t$ is the  Monte-Carlo (MC) time.
It also updates the histogram of the action, $h_t(E)$,
which is an auxiliary quantity used to estimate 
convergence. The steps of the algorithm are as follows~\cite{ZSTL}:
\begin{enumerate}
\item 
Make an initial guess for the entropy function at time ${t=0}$,
$\omega_0(E)$, using any available
prior knowledge, such as the results from a related system (e.g.
a smaller value of $N$ in our study). Set the histogram to zero:
$h_{t=0}(E)=0$. Pick any starting configuration.
\item 
Propose a new field configuration in an unbiased way, 
and accept it with probability:
\begin{equation}
{\rm Prob}(E\to E') = \min \left[ \, \exp{\left( \omega_t(E) -
\omega_t(E')\right)} , 1\,\right]\,,
\label{update_prob}
\end{equation}
where $E$ and $E'$ are respectively the action densities
of the original and proposed configurations. 
\item 
Repeat step (2) $N_{\rm hit}$ times for equilibration.
\item 
Let $E_t$ be the final value of the action density after step (3).
Update the entropy as follows:
\begin{equation}
\omega_t(E) \to \omega_{t+1}(E) = \omega_t(E) + \gamma
F_\delta(E,E_t)\,, \label{WL_update}
\end{equation}
where $\gamma>0$ and $F_\delta$ 
is a fixed, positive function,
which smears the update over a range
of action density of width $\sim \delta$ centered on $E_t$, 
and should be invariant under $E_t\leftrightarrow E$.
Possible choices for $F_\delta$ are discussed in Ref.~\cite{ZSTL}---we
use a simple Gaussian form
\begin{equation}
F_\delta(E,E_t) = e^{-(E-E_t)^2/\delta^2}. \label{F_delta}
\end{equation}
\item 
Update the histogram:
\begin{equation}
h_t(E) \to h_{t+1}(E) = h_t(E) + \delta(E-E_t).
\end{equation}
\item GoTo step (2).
\end{enumerate}

It is important to understand the meaning of the crucial step (4) :  if the
simulation has spent some time in the vicinity of
a particular value of $E$, then step (4) will increase $\omega_t$
in this region, and the update probability
(\ref{update_prob}) will favor motion to other regions
of $E$---this is how the WL algorithm encourages tunneling.

As we show in the following subsection, 
the WL algorithm converges in the sense that, for large enough $t$,
$\omega_t(E)-\omega(E)$ fluctuates around an $E$-independent constant. This constant drops out when one uses \Eq{OE} to calculate averages of physical observables, and so it is in this sense that 
\begin{equation}
\lim_{t\to\infty}\omega_t(E) = \omega(E)
\end{equation}
in the WL algorithm.

In Section~\ref{sec_tune},
we suggest practical ways to determine how large $t$ needs to be,
how to evaluate the errors in the estimate for $\omega(E)$,
and how to choose appropriate ranges for the parameters $\delta$
and $\gamma$.
In our implementation of the algorithm we restrict $E$ to
lie in the interval $E_{\rm min}\le E \le E_{\rm max}$, 
which is a subset of the full range of values $E$ can take. 
Apart from the need to begin with a configuration having $E$ inside
this range, the only change to the algorithm involve certain boundary effects that we discuss in 
Appendix~\ref{app_more}.

\subsection{Theoretical analysis of Wang-Landau reweighting}
\label{sec_WLA_theory}

A theoretical analysis of the original, `discrete' version, of the WL algorithm,
including some of its systematic errors, was given in Ref.~\cite{ZB}.
In this section we extend that analysis to the continuous WL algorithm
just described. 

Consider the probability distribution of the action density $E$ at MC-time
$t$ after step (3) of the algorithm has been completed.
Assuming that $N_{\rm hit}$ is large enough, this is
\begin{equation}
p_t(E) = \frac1{Z_t} \exp{\left(\omega(E)-\omega_t(E)\right)},
\label{p_def}
\end{equation}
where the normalization factor $Z_t$ ensures that $\int dE \ p_t(E) = 1$
(where here and in the following the $E$ integral implicitly runs
from $E_{\rm min}$ to $E_{\rm max}$).
This is the probability distribution from which $E_t$ is drawn. 
After updating $\omega_t(E)$ according to eq.~(\ref{WL_update}),
the function $p_t(E)$ changes as follows
\begin{equation}
p_t(E)\to p_{t+1}(E)=\frac1{Z_{t+1}}
\exp{\left[\omega(E)-\omega_t(E)-\gamma F_\delta(E,E_t)
\right]}\,,
\end{equation}
and a simple manipulation gives
\begin{equation}
\frac{p_{t+1}(E)}{p_t(E)} 
= 
\frac{\exp{\left[-\gamma F_\delta(E,E_t) \right]}}
{\left\<\exp{\left[-\gamma F_\delta(E,E_t) \right]}\right\>_t}.
\end{equation}
Here the average $\<,\>_t$ is with respect to the probability 
distribution at time $t$,
\begin{equation}
\left\<f(E)\right\>_t \equiv \int \, dE \, p_t(E)\, f(E).
\end{equation}

In order to understand the convergence properties
of the algorithm,
we need a measure of the closeness of the estimate $\omega_t(E)$ to
the true $\omega(E)$. When $\omega_t(E)=\omega(E)$, 
the probability distribution (\ref{p_def}) is flat, 
i.e. $p_t(E)=p_{\rm flat}(E)=1/\Delta E$.
Thus one possible measure of convergence is
\begin{equation}
\mu_t \equiv \int \, \frac{dE}{\Delta E} \, 
\log\left[\frac{p_t(E)}{p_{\rm flat}(E)}\right] 
= \int \, \frac{dE}{\Delta E} \, \log\left[\Delta E\, p_t(E)\right]\,.
\label{mut_def}
\end{equation}
This is adapted from the similar
discrete quantity used in Ref.~\cite{ZB}.
It is straightforward to see that $\mu_t\le 0$,\footnote{%
Given $1=\int dE \, p_t(E)=\int dE \, e^{log(p_t(E))}$,
use the identity $\int \left(\frac{dE}{\Delta E}\right)\, e^{f(E)} 
\ge \exp{\left\{ \int \left(\frac{dE}{\Delta E}\right) \, f(E)\right\}}$.}
with the upper bound saturated only when $p_t=p_{\rm flat}$.

We thus consider the change, $\Delta \mu_t = \mu_{t+1}-\mu_t$, 
between two adjacent time steps:
\begin{eqnarray}
\Delta \mu_t &=& \int ({dE}/{\Delta E})  \, \log[p_{t+1}(E)/p_t(E)] 
\\
&=& 
\int (dE/\Delta E) \left\{
-\gamma F_\delta(E,E_t) 
- \log \<\exp{\left[-\gamma F_\delta(E,E_t) 
\right]} \>_t \right\}\,.
\label{delta_mu_final}
\end{eqnarray}
Our choice of $F_\delta$, eq.~(\ref{F_delta}),
satisfies 
$\int dE\, F_\delta(E,E_t) = \delta \sqrt\pi$ for
all $E_{\rm min}\le E_t \le E_{\rm max}$ (This is true even taking into account any boundary effects -- see Appendix~\ref{app_more}). 
Thus
\begin{equation}
\Delta \mu_t = -\gamma\delta \sqrt{\pi}/\Delta E - 
\, \log \,
\<\exp{\left[-\gamma F_\delta(E,E_t) 
\right]} \>_t\,.
\label{deltaS}
\end{equation}
Since $\gamma>0$, the logarithm is always negative and 
$\Delta \mu_t$ is bounded from below 
\begin{equation}
\Delta \mu_t \ge -\gamma\delta \sqrt{\pi}/\Delta E\,. 
\end{equation}
Had this lower bound had been zero, then a monotonic convergence
of $\mu_t\to 0$ as $t\to\infty$ would have been possible.
A negative lower bound, however, suggests a more complicated
behavior involving fluctuations. In the rest of this section we describe the way these fluctuations emerge and quantify how they effect $\mu_t$, $\omega_t(E)$ and $p_t(E)$.

\subsubsection{$\Delta \mu_t$ as a function of $t$ and its ensemble average}
We begin by illustrating the possible values that $\Delta \mu_t$ can take. 
First consider the $\delta \to 0$ limit, in which,
assuming also that $\gamma\ll 1$, one finds\footnote{%
The assumption $\gamma \ll 1$ is valid for all our
calculations since we use $\gamma\simeq 10^{-4}-10^{-6}$.}
\begin{equation}
\Delta\mu_t = \gamma\delta\sqrt\pi 
\left(p_t(E_t)- p_{\rm flat}\right) + O(\gamma^2)\,.
\label{delatmu_size}
\end{equation}
Thus if $p_t(E_t)$ is above (below) 
the flat distribution value $1/\Delta E$,
then $\Delta \mu_t$ is positive (negative).
For large $t$, as we will see below, the generic size of
$|p_t-p_{\rm flat}|$ is $\sim \sqrt\gamma$, so that
$\Delta \mu_t$ then scales as $\gamma^{3/2}$.

Second, assume that at time $t$ one has $\omega_t(E)=\omega(E)$ so that
$p_t=p_{\rm flat}$ and $\mu_t$ takes its maximum value,
$\mu_t=0$. Since the algorithm updates the entropy,
$\omega_t(E) \to \omega_{t+1}(E)=\omega_t(E) + \gamma F_\delta(E,E_t)$,
$\Delta \mu_t$ must be negative. A simple calculation gives
\begin{equation}
\Delta \mu_t = -\frac{\gamma^2 \delta \sqrt{\pi} }
{\Delta E\,\sqrt{8}} \left( 1 - \sqrt{2\pi} \frac{\delta }{\Delta E}\right) +  
O(\gamma^3) \,.
\label{negative_Deltamu}
\end{equation}
The size of this rather special step is parametrically smaller than
the generic $O(\gamma^{3/2})$ of the first example.
Note that the exact density of states is not a ``fixed point'' of the
algorithm, which may be surprising at first glance, but is in fact an
essential feature of the algorithm. It ensures that the
simulation explores all values of $E$ in the desired range.

To get a more precise measure of how  $\Delta\mu_t$ behaves, we calculate its
expectation value  averaged over an ensemble of simulations 
all starting with the same $\omega_t(E)$.
The result is
\begin{eqnarray}
\<\Delta \mu_t \> &\equiv& \int dE_t \, \Delta \mu_t \, p_t(E_t) 
\\
&=& -\frac{\gamma\delta \sqrt{\pi}}{\Delta E} 
- \int dE_t \, \, p_t(E_t) \, \log \left[
1-\left\<1 - \exp{\left(-\gamma F_\delta(E,E_t) 
\right)}\right\>_t\right].
\end{eqnarray}
(Here the internal average is over $E$.) 
Using the identity $\log (1-x)^{-1} > x$ we find
\begin{equation}
\<\Delta \mu_t\> > -\frac{\gamma \delta \sqrt{\pi}}{\Delta E} + 
\int dE_1 \, dE_2 \,\, p_t(E_1) \,\, p_t(E_2) \, 
\left\{ 1- \exp{\left[ -\gamma
F_\delta(E_1,E_2) 
\right]} \right\}\,. 
\label{deltaS_vev}
\end{equation}
The kernel in the curly braces is positive semi-definite,
but decreases rapidly towards zero for $|E_1-E_2|\gg \delta$.
It is also symmetric under $E_1\leftrightarrow E_2$. 
Consequently, the double integral on the r.h.s. of (\ref{deltaS_vev}) 
provides a definition of a (smeared) inner product of $p_t(E)$ with itself.

\subsubsection{Relation of $\Delta \mu_t$ to $p_t(E)$}

To make use of \Eq{deltaS_vev} we must evaluate the
second term on the r.h.s. and relate it back to $\mu_t$.
For that purpose we first use the kernel and  define the 
following squared ``distance'' 
between two probability distributions
(\ref{deltaS_vev}):
\begin{equation}
||p_a-p_b||^2 \equiv 
\frac{\int dE_1 \, dE_2 \,
  [p_a(E_1)-p_b(E_1)]\, 
\left\{ 1- \exp{\left[ -\gamma
F_\delta(E_1,E_2) 
\right]}\right\}
  [p_a(E_2)-p_b(E_2)]} 
{\int (dE_1/\Delta E) \, (dE_2/\Delta E) \, 
\left\{ 1- \exp{\left[ -\gamma
F_\delta(E_1,E_2) 
\right]} \right\} }\,. 
\label{norm_def}
\end{equation}
The normalization is chosen so that $||p_{\rm flat}||^2=1$.
Equation (\ref{norm_def}) is a generalization of the standard 
Euclidean distance used in \cite{ZB}. In fact, if one
takes $\delta\to 0$, the kernel becomes proportional to
$\delta (E_1-E_2)$, and one obtains (the continuous $E$ version of)
the Euclidean distance:
\begin{equation}
\lim_{\delta\to 0} ||p_a-p_b||^2 = \int \left( \frac{dE}{\Delta E} \right) \,
  \left(\frac{ p_a(E)-p_b(E)}{p_{\rm flat}}\right)^2\,.
\label{delta_to_0}
\end{equation}
For $\delta > 0$, the kernel gives different 
weights to different Fourier components (in $E$-space) of
$(p_a(E)-p_b(E))$: wavelengths larger than $\delta$
are included with full weight, with the weight decreasing to zero as
the wavelength itself decreases to zero. As a result UV differences are filtered out.
Indeed this kernel is a natural integration measure for our purposes because
the WL algorithm only makes changes to $\omega_t$ which have wavelengths of
$O(\delta)$ or longer.

We can use the distance $||p_t-p_{\rm flat}||$
as an measure of the approach of
$p_t$ to $p_{\rm flat}$. To evaluate this distance we need to
calculate the integral in the denominator of
\Eq{norm_def}
\begin{equation}
\int (d E_2/\Delta E) 
\left\{ 1- \exp{\left[ -\gamma
F_\delta(E_1,E_2) 
\right]} \right\} 
\equiv [1 - c]  \gamma \delta \sqrt\pi/\Delta E \,.
\label{c_def}
\end{equation}
The constant $c$ obeys $0< c \le 1$, and for small $\gamma$ is 
\begin{equation}
c = \frac{\gamma}{\sqrt8} + O(\gamma^2) \,.
\end{equation}
It is independent of $E_1$ up to boundary effects of
$O(\gamma^2 \delta/\Delta E)$. Ignoring these numerically very small effects,
it is straightforward to show that
\begin{equation}
||p_t-p_{\rm flat}||^2 = ||p_t||^2 - 1 \,. 
\label{pt_minus_pflat}
\end{equation}
Combining Eqs.~(\ref{deltaS_vev}), (\ref{pt_minus_pflat})
and (\ref{c_def}), we find
\begin{equation}
\<\Delta \mu_t \> > \frac{\gamma\delta \sqrt{\pi}(1-c)}{\Delta E} \, 
\left(||p_t-p_{\rm flat}||^2-R^2\right). 
\label{Delta_mu_av}
\end{equation}
with $R^2 \equiv c/(1-c) \simeq \gamma/\sqrt{8}$. 

From this it follows that
\begin{itemize}
\item
If $||p_t-p_{\rm flat}|| >  R$ then 
$\<\Delta \mu_t \> > 0$ and the simulation will, on average, 
move towards the desired point $p_t=p_{\rm flat}$, at which
$\omega_t(E)=\omega(E)$.
\item If $||p_t-p_{\rm flat}|| < R$ then the lower bound on
$\<\Delta \mu_t\>$ is negative and the simulation can move
both towards and away from $p_t=p_{\rm flat}$.
\end{itemize}
Finally, note that when
$|(p_t-p_{\rm flat})/p_{\rm flat}| \ll 1$, one can show 
from the definition of $\mu_t$ [\Eq{mut_def}] that
\begin{equation}
\mu_t \approx -\frac12
\int \left( \frac{dE}{\Delta E} \right) \,
  \left(\frac{ p_t(E)-p_{\rm flat}(E)}{p_{\rm flat}}\right)^2
\approx -\frac12 ||p_t-p_{\rm flat}||^2
\,.
\label{mu_t_vs_distance}
\end{equation}
The first approximate equality assumes that the fluctuations of 
$(p_t(E)-p_{\rm flat})$ are small (which is a good approximation
at large $t$, as we will see shortly).
The second approximate equality assumes that the corrections
to the $\delta\to 0$ limit, \Eq{delta_to_0}, are small,
and is thus only an order of magnitude approximation.


\subsubsection{Behavior of $p_t$ as a function of $t$ and estimating fluctuations}
\label{fluc}

\begin{figure}[htb]
\includegraphics[width=15cm]{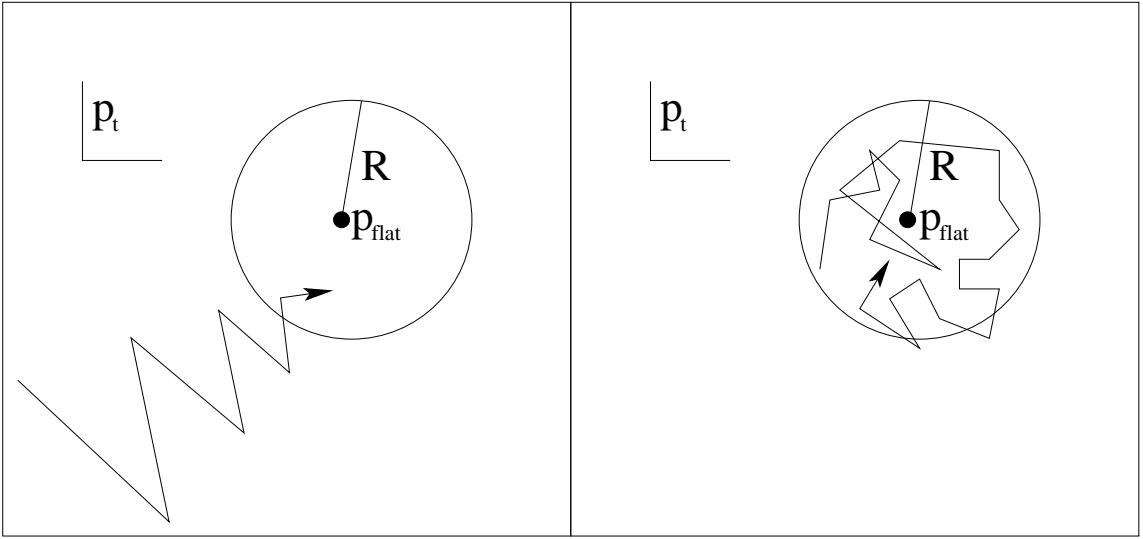}
\caption{Pictorial representation of how the 
Monte-Carlo time history of the  WL algorithm looks in $p_t(E)$ space (see
text). \underline{Left panel:} Initial stage - convergence of $p_t(E)$
towards a `ball' of radius $R$ around $p_{\rm flat}$. The step-size 
gets smaller as the algorithm approaches the ball. \underline{Right
panel:} Second stage - fluctuations within the ball. 
If the fluctuations drive $p_t$ outside of the ball, 
it is driven back inside by the type of motion in the left panel.  
\label{ball}
}
\end{figure}

Putting together the above ingredients, the following picture emerges.
Consider the infinite dimensional 
space of probability distributions $p_t(E)$,
with distances defined by the Euclidean metric \Eq{norm_def}. 
Let the origin be at $p_t(E)=p_{\rm flat}$, and
denote the radial coordinate in this space, $||p_t-p_{\rm flat}||$, by $r$.
A crucial role is then played by a ball of radius 
$R\approx\sqrt{\gamma/\sqrt8}$ centered at the origin.
The results above imply that if $p_t(E)$ lies outside this ball,
then the simulation will perform a directed random
walk towards the ball, with steps in $\mu_t$
[and thus, from \Eq{mu_t_vs_distance}, also in $r^2$]
of average size proportional to $(r^2-R^2)\times\gamma\delta/\Delta E$.
Since the steps get, on average, smaller as one approaches
the ball, the approach to its surface is exponentially slowed.
Individual steps, however, do not shrink to zero, so one
will eventually end up inside the ball.
Once inside,
\Eq{Delta_mu_av} only gives a lower bound on $\<\Delta\mu_t\>$,
so we do not know its sign. The simulation may
move throughout the ball, or it may cluster near the surface.
Combining Eqs.~(\ref{mu_t_vs_distance}) and (\ref{negative_Deltamu}), 
one finds that the typical step size is
$\Delta r\approx\gamma\delta$ 
and thus much smaller than
the size of the ball $R\sim \gamma^{1/2}$.
One possible behavior is illustrated in Fig.~\ref{ball}.

It is clear from the foregoing that only in the second stage,
when the simulation has reached the ball, can one use 
$\omega_t(E)$ as an estimate of $\omega(E)$. 
We now calculate the size of the fluctuations in this estimate.
This requires that we remove the overall uniform growth of
$\omega_t$ which occurs because of the addition of $\gamma F_\delta(E,E_t)$
with a uniform distribution of $E_t$.
To do so we write:
\begin{equation}
\omega_t(E) - \omega(E) = C(t) + \Delta \omega_t(E)\,,\qquad
{\rm with}\qquad
\int dE\, \Delta\omega_t(E) = 0\,.
\label{ominball}
\end{equation}
The $E$-independent quantity $C(t)$ is determined by the normalization
condition on $\Delta\omega_t$, and is a linear function of $t$
with slope $\delta\gamma\sqrt\pi/\Delta E$.
$\Delta\omega_t(E)$ contains the physically relevant fluctuations, since
$C(t)$ makes no contribution to observables.

Once we are inside the ball we have
$||p_t-p_{\rm flat}||\stackrel{<}{_\sim} R \approx \sqrt{\gamma/\sqrt8} \ll 1$.
Our task is to use the definition of $p_t$ in  \Eq{p_def}
to convert this into a result for the fluctuations in $\omega_t$. 
To do so, we assume that once in the ball, the proximity of 
$p_t(E)$ to $p_{\rm flat}(E)$ occurs not just on average 
(as the smallness of $||p_t-p_{\rm flat}||$ implies) 
but also for each $E$ separately. 
Then we have that
\begin{equation}
\frac{p_t(E)-p_{\rm flat}}{p_{\rm flat}}
\approx 
\log\left(\frac{p_t(E)}{p_{\rm flat}}\right)
=-\Delta\omega_t(E) + O\left( \Delta\omega_t(E) \right)^2
\,.
\label{deltap_vs_deltaomega}
\end{equation}
The last step follows by expanding $Z_t$ in $\Delta\omega_t(E)$.
Inserting the result (\ref{deltap_vs_deltaomega}) into \Eq{norm_def} 
we find the desired relation:
\begin{equation}
||p_{\rm flat}-p_t||^2  \approx
\frac{\int dE_1 dE_2
 ( \omega_t(E_1)-\omega(E_1)) \left( 1-\exp{\left[-\gamma F_\delta(E_1,E_2)
\right]} \right) (\omega_t(E_2) - \omega(E_2))}
{\int dE_1 dE_2 \left(
1-\exp{\left[-\gamma F_\delta(E_1,E_2)
\right]} \right) }
\equiv 
(\Delta \omega )^2\,.
\end{equation}
Thus, once in the ball, the fluctuations in $\omega_t$ are the 
same as those in $p_t$.
Such a ``filtered'' measure of fluctuations
is sufficient, because the update of $\omega_t$ does not introduce UV noise.
We conclude that $\Delta\omega \approx R \approx \sqrt{\gamma/\sqrt8}$.
This is the same parametric behavior as in the discrete WL algorithm~\cite{ZB}.

An important issue for the practical application of our
 variant of the WL algorithm is the detailed
 nature of the fluctuations $\omega(E)-\omega_t(E)$.
 In particular,
 do they average to zero {\em for each $E$} once one is inside the ball?
 The previous analysis does not directly address this question.
 We consider it very plausible, however, that the answer is positive.
 This is because the algorithm is designed to smooth out nonuniformities
 in $\omega(E)-\omega_t(E)$, although it does so with
 some ``overshoot'' which leads to the fluctuations.
 It would be interesting to extend the analysis of the algorithm to
 include such non-equilibrium effects. For the present, however,
 we assume that $\omega(E)-\omega_t(E)$ fluctuates symmetrically
 about zero for each $E$. 

We end this subsection by estimating the parametric dependence of
fluctuations in the histogram $h_t(E)$, at least in certain
limits.
The entropy is related to the histogram by a Gaussian transform,
\begin{equation}
\omega_t(E) -\omega_0(E) = 
\gamma \int dE' \, h_t(E') \, e^{-(E-E')^2/\delta^2}\,,
\label{omega_from_h}
\end{equation}
where $\omega_0$ is the initial guess.
We will use a shorthand notation for this transform
and its inverse:
\begin{equation}
(\omega_t-\omega_0) = \gamma {\cal G}(h_t)\,,\qquad
{\cal G}^{-1}(\omega_t-\omega_0) = \gamma \,h_t. \label{Ginv}
\end{equation}
Using \Eq{ominball}, we can write
\begin{equation}
\omega_t(E)-\omega_0(E) = C(t) \,+\, 
\left(\omega(E)-\omega_0(E)\right) +  \sqrt\gamma f_t(E)\,.
\label{omega_t_form}
\end{equation}
For large $t$, when $p_t$ is in the ball, 
$f_t=\Delta\omega_t/\sqrt\gamma$ fluctuates around zero with
an amplitude of $O(1)$.
Substituting \Eq{omega_t_form} into \Eq{Ginv} we obtain
\begin{equation}
h_t = \frac{C(t)}{\gamma\delta\sqrt\pi} \,+\, 
{\cal G}^{-1}\left(\omega-\omega_0\right)\,/\,\gamma
+ {\cal G}^{-1} (f_t) / \sqrt\gamma\,,
\label{h_t_form}
\end{equation}
where we use that result that an inverse Gaussian 
transform of a constant is a constant.

The subsequent analysis depends on the relative size of
the second and third terms in (\ref{h_t_form}),
and thus on the accuracy of the initial guess $\omega_0(E)$. 
One extreme case is a poor guess, $\omega_0=0$. In this case
the second term dominates over the third (at least for small enough
$\gamma$) which means that if one evaluates the variance in $h_t$,
\begin{eqnarray}
\delta h_t &\equiv&
\sqrt{\int ({dE}/{\Delta E}) 
\left(h_t(E)-\overline{h_t}\right)^2}\,,\\
\overline{h_t} &\equiv& \int \, (dE/\Delta E) \, h_t(E).
\label{def_deltah}
\end{eqnarray}
it will have a $t$-independent contribution proportional to $1/\gamma$.

The other extreme is when one starts with a very good guess, 
$\omega_0(E)\approx\omega(E)$, so that the fluctuation
term in (\ref{omega_t_form}) dominates over the second term
on the r.h.s.. 
If so, then we expect a $t$-dependent contribution to $\delta h_t$
that scales with $1/\sqrt{\gamma}$.
Presumably if the second and third terms compete, the
scaling will lie somewhere between these two limiting cases.
This appears to be the situation in many of our simulations.

The result (\ref{omega_from_h}) tells us nothing, however, about the
UV fluctuations in $h_t$, since these are filtered out
by the Gaussian transform. As discussed further below, 
we expect that this UV noise increases with $t$. 
In practice it is a small contribution in our simulations.

\subsection{The effect of a non-equilibrated Wang-Landau simulation}
\label{Nhitsmall}

We close this section by stressing that the analysis just presented is
predicated on letting the simulation equilibrate after an update to
$\omega_t$ is performed. While this equilibration is guaranteed if we
let $N_{\rm hit}\to \infty$, most of our runs were done with
$N_{\rm hit}=1$.  This means that the analysis above does not
directly apply to such simulations---$p_t(E)$ is changing
after each update,
\begin{equation}
p_t(E) \to p_{t+1}(E) = p_t(E) + O(\gamma),
\end{equation}
so exact equilibration cannot occur.
Nonetheless, when $\gamma\ll1$ and 
$p_t(E)\simeq p_{t+1}(E)$,
approximate equilibration is possible.
Thus we think it is plausible that the analysis just
given remains applicable given $\gamma$ is small enough. 
We have checked this in practice
by doing runs with $N_{\rm hit}\gg 1$ and seeing that
the results are unchanged within errors. An example
is shown below.

\section{Implementing and Tuning the Wang-Landau Algorithm}
\label{sec_tune}

In this section we describe how we implement the WL algorithm
in practice, how we use it to estimate $\omega(E)$ and derived
quantities, and suggest
criteria for tuning $\gamma, \delta$ and $N_{\rm hit}$.

That tuning is necessary is apparent from the
analysis of the previous Section.
Particularly crucial is the tuning of $\gamma$,
which involves a balance between two competing effects.
On the one hand, $\gamma$ controls the speed with which
the algorithm explores values of action density. If the
simulation has spent some time in the vicinity of
a particular value of $E$, then $\omega_t$ will be
increased in this region, and the update probability
(\ref{update_prob}) will favor motion to other regions
of $E$. The rate of build-up of $\omega_t$
is proportional to $\gamma$, so the rate of motion through
``$E$-space'' will increase with increasing $\gamma$.
On the other hand, by reducing $\gamma$ one 
reduces the fluctuations in $\omega_t$
(since $\Delta\omega\propto\sqrt\gamma$),
and correspondingly reduces statistical errors in quantities derived
from $\omega$.

\subsection{Algorithm structure and tuning $\gamma$}
\label{gamma_fix}

For our application we can restrict the range of $E$ to
$[E_{\rm min},E_{\rm max}] \subset [-1,1]$, since we are only interested
in the transition region. The range should be large enough
that the errors in the quantities of interest due to
this truncation are much smaller than those from statistics.
The appropriate range in our case can be read off from the
hysteresis curves of Fig.~\ref{hyst_map}. One must cover
the transition region and add a conservative cushion on
each side, and we typically use $E\in [0.1,0.7]$ (see below
for all our parameter choices). Working with less than a
third of the full range $[-1,1]$ saves considerable
computation time.

Having made the choice of range, the algorithm proceeds in two
stages.
\subsubsection{Initial stage:}

During this stage the simulation makes a directed
random walk towards the ``ball'' in $p_t$ space of radius $R$
centered on the desired flat distribution.
The algorithm explores the chosen range of $E$ and transforms
the starting guess $\omega_0(E)$ into a reliable estimate
of the actual entropy.

The histogram $h_t(E)$ is a useful
monitor of progress during this stage.
At the beginning, it will build up non-uniformly because the
guess for the entropy function is imperfect, but by the end 
the histogram should be growing uniformly in $E$.
The variance of the histogram, $\delta h_t$, will grow
from its initial value of zero 
and then approximately saturate. 
This saturation marks the end of the initial stage.

\begin{figure}[htb]
\includegraphics[width=10cm,angle=-90]{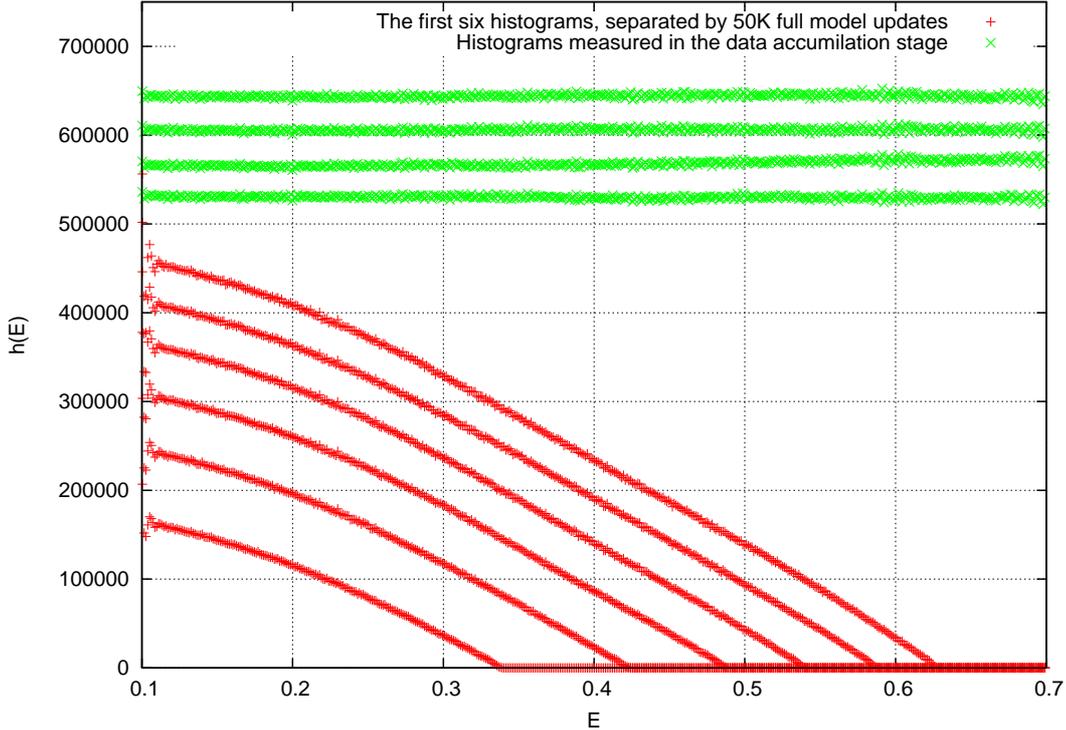}
\caption{Comparison of histograms $h_t(E)$ in the initial stage 
with $\omega_0=0$ (red pluses) and 
the data accumulation stage for a good initial guess for $\omega_0$ 
(green crosses). Results using WLR for
$SU(20)$ with $\gamma=10^{-4}$, $\delta=0.005$.
The values of $t$
that correspond to the bottom six histograms 
$t=[5,10,15,20,25,30]\times 10^4$ full updates of the model
(corresponding to values of 25--150 on the ``MC-time'' axis
of the next figure). 
The top four histograms were obtained after $[50,55,60,65]\times
10^4$ full updates.
The definition of a ``full update'' in given in Appendix~\ref{app_alg}.
\label{h_initial_blind}
}
\end{figure}
As discussed in the previous section, the value at
which it saturates depends on the accuracy of the
initial guess. We illustrate in Fig.~\ref{h_initial_blind} 
what happens with both a poor and a good guess.
In the former case (data represented by [red] pluses) 
we start with no information on the entropy, i.e. $\omega_0=0$. 
The histogram first increases for small $E$, where $\omega(E)$ is
large. When $\omega_t(E)\simeq \omega(E)$ for these values of $E$,
the WL random walk gradually starts exploring larger values of $E$
which have lower $\omega(E)$. Eventually (not shown) 
the whole range is covered,
and the histogram grows uniformly, while maintaining in its shape the
``memory'' of the initial $\omega_0$.
This shape is the second term in \Eq{h_t_form}.

The case of a good guess is shown by the [green] crosses. Here we show
only the histograms starting after $55\times 10^5$ updates, so
as to avoid cluttering the figure.
Earlier histograms are similarly horizontal.
For these simulations the third (fluctuation)
term in \Eq{h_t_form} may dominate over the second.

\begin{figure}[htb]
\includegraphics[width=10cm,angle=-90]{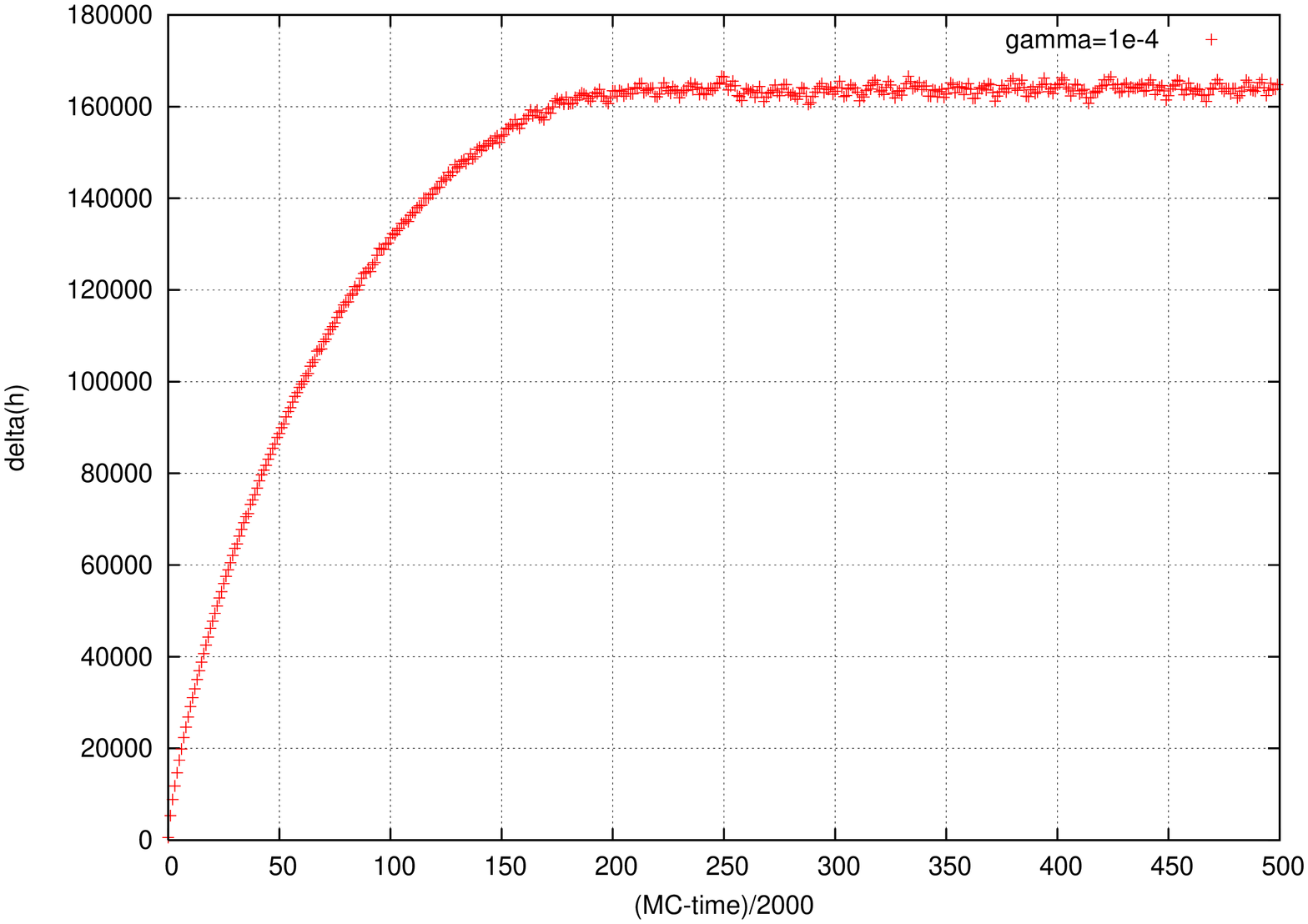}
\caption{The average histogram variance, $\delta h_t$, for the 
$\omega_0=0$ data
presented in Fig.~\ref{h_initial_blind}.
\label{delta_h_initial_blind}
}
\end{figure}

The behavior of $\delta h_t$ for the case of the poor guess is shown in
Fig.~\ref{delta_h_initial_blind}.
The stage of rapid growth ends when $\delta h_t$
saturates to a nearly constant function of $t$.\footnote{%
As noted in the previous section, we expect the amplitude
at saturation to scale with $1/\gamma$, 
although we have not checked this in this case.}
There is a small but noticeable residual growth in
$\delta h_t$ which is due, we think, to UV fluctuations in the
histogram. 
As discussed in Sec.~\ref{sec_WLA_theory}, these fluctuations 
are not suppressed by the WL algorithm, 
and we expect them to be Gaussian with a contribution
to $\delta h_t$ growing like $\sqrt{h_t}$.

\begin{figure}[htb]
\includegraphics[width=10cm,angle=-90]{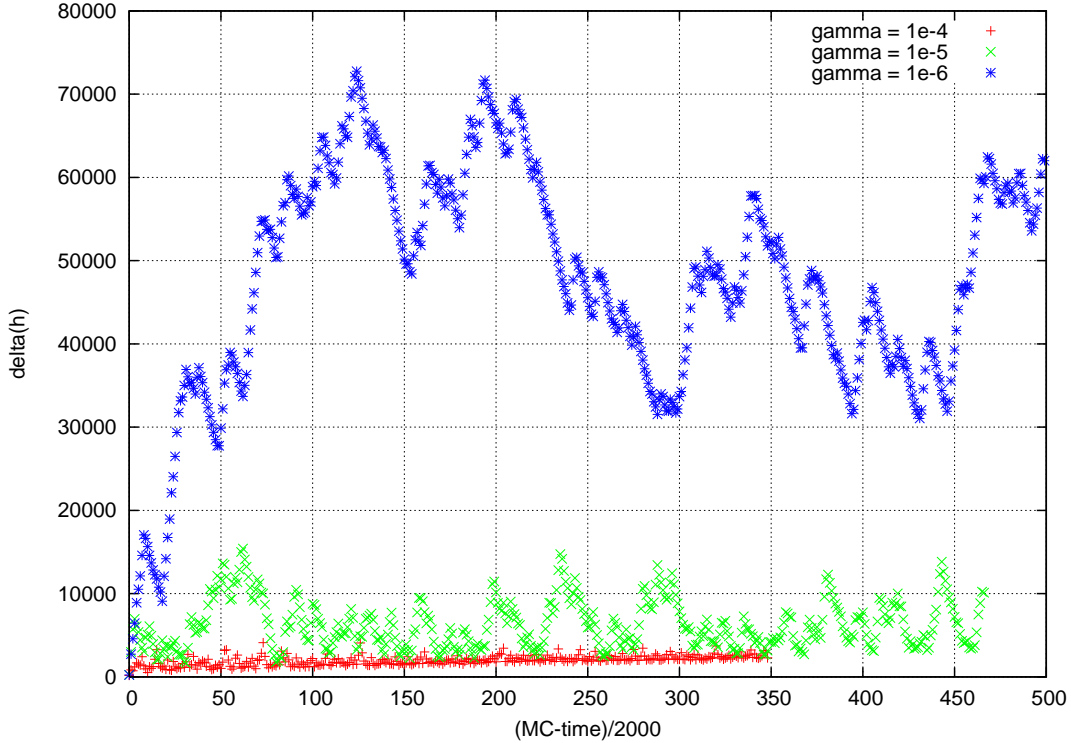}
\caption{The standard deviation in the average histogram, $\delta h_t$,
versus MC-time,
for $SU(20)$ and $\gamma = 10^{-4},10^{-5},10^{-6}$
(shown by [red] pluses, [green] crosses and
[blue] stars, respectively). This data uses
a good initial guess $\omega_0(E)$ and so has a
very short initial stage. Thus $\delta h_t$ is
mostly or completely saturated. Results are for $\delta=0.005$
and $N_{\rm hit}=1$. The corresponding plot for $N_{\rm hit}=20$ and
 $\gamma=10^{-5}$ is similar.
\label{saturation_comp_gamma}
}
\end{figure}

Once one has obtained a good estimate of $\omega(E)$ for
one value of $N$, one can scale it with $N_{\rm dof}\propto N^2$ 
to use as a guess for a different value of $N$, or reuse it
for the same $N$ with a different $\gamma$, $\delta$ etc..
With a good guess the initial stage is shorter,\footnote{%
Ref.~\cite{ZSTL} reports that
introducing an update to $\omega_t(E)$ which is applied simultaneously
to all values of $E$ can also reduce the computational cost of this initial
stage by an order of magnitude. We did not test this extensively.}
and, according to Section~\ref{fluc}, 
the value of $\delta h_t$ at saturation 
should scale more like $1/\sqrt{\gamma}$ than like $1/\gamma$.
An example of this situation is shown in Fig.~\ref{saturation_comp_gamma} where
it appears that most, if not all, of the data is in the ``saturation regime'',
although it is hard to pinpoint exactly the beginning of this 
regime because of the fluctuations.
Note that the $\gamma=10^{-4}$ data correspond to the ``good guess''
histograms in Fig.~\ref{h_initial_blind} above.
The figure shows clearly that the saturated $\delta h_t$ grows
with decreasing $\gamma$. The saturated values are approximately
$5\times 10^{-4}$, $8 \times 10^{-3}$ and $ 2.5\times 10^{-3}$,
which are roughly consistent with the expected $\gamma$-scaling.

Finally we note that we found it useful to 
experiment during this initial stage with values for $\gamma$,
and determine a lower bound such that the range
$[E_{\rm min},E_{\rm max}]$ can be explored repeatedly
with the available computational resources.

\subsubsection{Data accumulation stage:}

The simulation is now fluctuating around the actual $\omega(E)$ (it is in the ball -- see Section~\ref{fluc}).
We propose that one perform $N_{\rm meas}$
measurements of $\omega_t(E)$ separated by a fixed number 
updates. In our case of a first-order transition, the gap between
measurements should ideally include, on average, several 
tunneling events.\footnote{%
We define a tunneling event as motion from $E_{\rm min}$ to
$E_{\rm max}$ and back again. This is a conservative definition
since a tunneling can occur without motion all the way to
the edges of the range.}
The average of these measurements provides an estimate for
$\omega(E)$ (up to an overall irrelevant constant).
The deviation of this estimate from the true entropy
will then scale as $\sqrt{\gamma/N_{\rm meas}}$. 

For derived quantities such as the specific heat (\ref{spec_heat}),
we propose calculating the errors using the jack-knife or similar method
applied to the set of $N_{\rm meas}$ measurements of $\omega_t(E)$.
This has the advantage of automatically taking into account
correlations in cases where we have two few tunneling events between
measurements. We expect the errors in derived quantities
to also scale as $\sqrt{\gamma/N_{\rm meas}}$. 

\begin{figure}[htb]
\includegraphics[width=10cm,angle=-90]{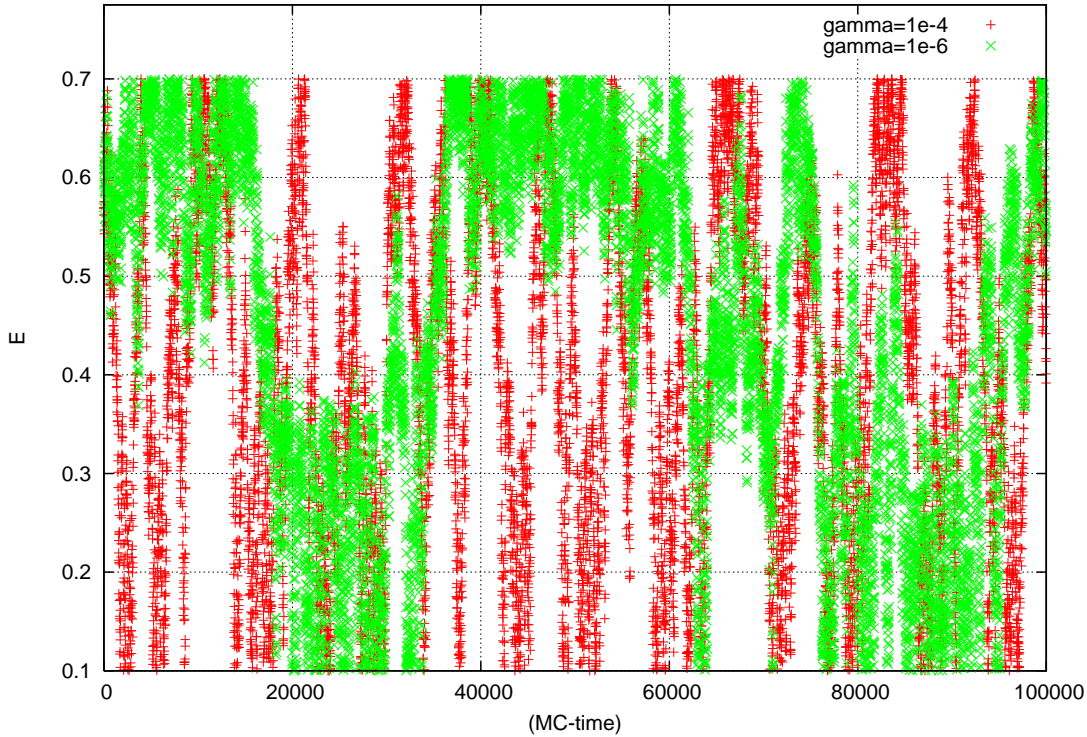}
\caption{The MC-time history of the action density $E$ for $SU(20)$
with $\gamma = 10^{-4}$ (red pluses) and $\gamma=10^{-6}$ (green
crosses). In both cases $\delta=0.005$ and $N_{\rm hit}=1$.
\label{history_E_su20}
}
\end{figure}

We show an example of the behavior of $E$ during this data accumulation stage in Fig.~\ref{history_E_su20}. 
The runs are the same as for Fig.~\ref{saturation_comp_gamma},
except that we show only the smallest and largest
values of $\gamma$ for the sake of clarity.
Tunneling is clearly seen,\footnote{%
It is important to keep in mind that these tunneling histories inevitably
look different from those in canonical simulations running at or near
the transition coupling. In the latter the fluctuations in each phase are
over a very limited range of $E$, while in the WL algorithm the simulation
must, by construction, move out to the boundaries of the $E$-range so
that all values of $E$ are equally populated.}
with a frequency that decreases
with decreasing $\gamma$.

This time history allows one to understand the large fluctuations
in $\delta h_t$ seen for small $\gamma$ in Fig.~\ref{saturation_comp_gamma}.
Before a tunneling event takes place, the histogram grows only for
low (high) values of $E$, and consequently $\delta h_t$ grows. After
the tunneling, the previously unvisited high (low) range of $E$ is
explored and $\delta h_t$ drops. Thus a tunneling event is manifest in
the MC-time history of $\delta h_t$ as a peak.
Indeed we have confirmed that the peaks in Fig.~\ref{saturation_comp_gamma}. 
peaks coincide with tunneling events
seen in the time histories of $E$ shown in Fig.~\ref{history_E_su20}.

The data accumulation stage can also be used to further
tune the value of $\gamma$. Decreasing $\gamma$ reduces
errors, but also, for fixed computer time, decreases tunneling
rates and thus $N_{\rm meas}$. One should choose $\gamma$ to optimize
the error in the derived quantities of most interest, which, as
we have noted, are expected to scale like $\sqrt{\gamma/N_{\rm meas}}$.

\begin{figure}[htb]
\includegraphics[width=10cm,angle=-90]{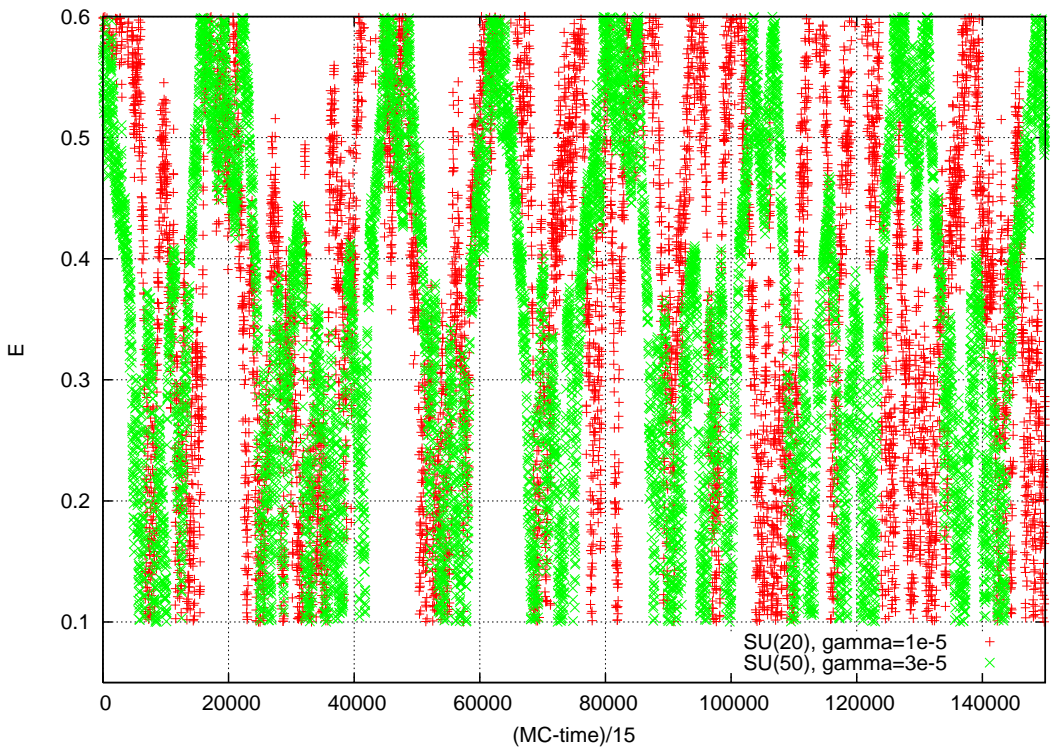}
\caption{The MC-time history of the action density $E$ using the WL
algorithm. \underline{Red pluses:} $SU(20)$ with
$\gamma=10^{-5}$. \underline{Green crosses:} $SU(50)$ with
$\gamma=3\times 10^{-5}$. In both cases $\delta=0.005$ and
$N_{\rm hit=1}$.
\label{su20_su50_tun_comp}
}
\end{figure}

One must also determine how to scale an optimized $\gamma$ between
different values of $N$.
One criterion is to maintain the same tunneling rate.
Since $\omega(E)$ is an extensive quantity scaling like $N_{\rm dof}$, 
we expect that $\gamma$ must be scaled similarly if it is to lead
to a similar rate of motion through $E$-space, and in particular
to the same tunneling rate. 
We have found that such a scaling rule works reasonably well
in practice. As an example, we show
in Fig.~\ref{su20_su50_tun_comp} the comparison of
the time-history for $SU(20)$ with $\gamma=10^{-5}$ and $SU(50)$
with $\gamma=3\times 10^{-5}$.
The ratio of the $\gamma$'s is $3$ while the ratio in
the number of degrees of freedom is 
$(50/20)^2\simeq 6$. Thus they are of the same order of magnitude and we expect the tunneling rate
for $SU(50)$ to be similar to that for $SU(20)$.
As the Figure shows this is approximately true.
This should be contrasted with standard MC simulations in 
which the tunneling rate for $SU(50)$ is exponentially smaller,
reduced in the present case by about a factor of $500$ compared
to $SU(20)$. This is a striking example of the efficacy of
the WL algorithm at overcoming the suppression of tunneling events.

\subsection{Tuning $\delta$}
\label{tuning_delta}

The parameter $\delta$ determines the width of the
smearing function $\gamma F_\delta(E, E_t)$ that is added to $\omega_t(E)$.
Since the area under $\gamma F_\delta$ is proportional to
$\gamma\times \delta$, it is this product that determines
how fast the simulation moves through $E$-space. 
Indeed this product enters in the bound on the steps
in $\Delta\mu_t$, \Eq{Delta_mu_av}.
By contrast, the size of fluctuations in $\omega_t$, and thus
in derived quantities, depends only on $\gamma$ and not
on $\delta$ (since $R^2\approx \gamma/\sqrt8$).
In light of this one wants to make $\delta$ as large as possible
before tuning $\gamma$. 

The upper limit on $\delta$ is set by different considerations.
As $\delta$ increases, the resolution with which one obtains
$\omega(E)$ is decreased, and it must not approach the width
of the region which makes the important contributions to
observables like the specific heat. Thus we propose that one must
keep $\delta \ll \sigma$, with $\sigma$ the
width in $E$ of each branch of
the canonical distribution $P_C(E)$,
\begin{equation}
P_C(E) \sim \exp \left(\omega(E) + 12N^2 b_t E \right)\,,
\label{p_c_def}
\end{equation}
in the vicinity of the transition coupling.
This guarantees that the integral in
\Eq{OE} can be evaluated accurately. 
We note that $\sigma$ can be estimated with a
standard MC simulation.

In practice we choose a value, $\delta=0.005$, 
which clearly satisfies $\delta\ll \sigma$ 
(see Fig.~\ref{P_final} below)
and do not undertake extensive investigations of the sensitivity
to this choice. 

\subsection{Tuning $N_{\rm hit}$}
\label{tuning_Nhit}

Finally, we discuss the tuning of $N_{\rm hit}$, which we recall
is the number of updates one does with a given $\omega_t(E)$ before
updating to $\omega_{t+1}(E)$. 
To reduce computational effort,
one wants to choose $N_{\rm hit}$
as small as possible. 
This, however, can introduce a sizable systematic error since 
the convergence of the WL algorithm is formally only guaranteed
if $N_{\rm hit}\to \infty$.
The lower limit $N_{\rm hit}$ depends on $\gamma$.
This is because for large values of $\gamma$, the update of
\Eq{WL_update} is very abrupt and the system will require more
hits to equilibrate into the new distribution $p_{t+1}(E)$. 
Correspondingly, for the very small values of $\gamma$ that we use,
the system may be able to equilibrate even with $N_{\rm hit}=1$.
In fact we use this value for most of our runs.

Lacking a firm theoretical foundation, it is clearly important
to do numerical checks of the dependence on $N_{\rm hit}$.
What we find (as will be shown below) is that $N_{\rm hit}=1$
is acceptable (i.e. gives the same results as with larger values)
if $\gamma$ is small enough. A possible explanation for this is that the 
number of effective hits between updates of the entropy is larger than $N_{\rm hit}$.
This is because the typical change in $E$ in an individual update
is, in our simulations, an order of magnitude smaller than $\delta$.
Thus the system performs a random walk ``inside'' the Gaussian $F_\delta$.
So, in an approximate sense, the simulations are
being done with effective values of $N_{\rm hit}$ and
$\gamma$ that are two orders of magnitude larger than the assigned values.

We conclude this section by stressing that this systematic
error should be estimated explicitly.
This can be done by comparing results
for derived quantities to those obtained using standard
MC simulations at values of $b$ away from the transition
(so that the latter are reliable), 
and/or by checking the sensitivity of 
the results obtained with WLR to changes in
$\gamma$ and $N_{\rm hit}$.
Details of such checks will be described at the end of the next section.
\section{Results}
\label{sec_res}

We have undertaken long runs with $N=20-50$ using the
parameters listed in 
Tables~\ref{tab_res_su20}--\ref{tab_res_su50}. 
(The parameter we denote by $N_{\rm
bin}$ is discussed in Appendix~\ref{app_more}.)
In all cases the
measurements were separated by $10000$ full updates of the model 
(for a definition of a full update see Appendix~\ref{app_alg}),
except for the data in the last two rows of Table~\ref{tab_res_su20},
where the separation was by $100000$ full updates.
Thus for each choice of $N$ and algorithm parameters, we perform
in total $(1-3)\times 10^6$ full updates.

\begin{figure}[t]
\includegraphics[width=15cm]{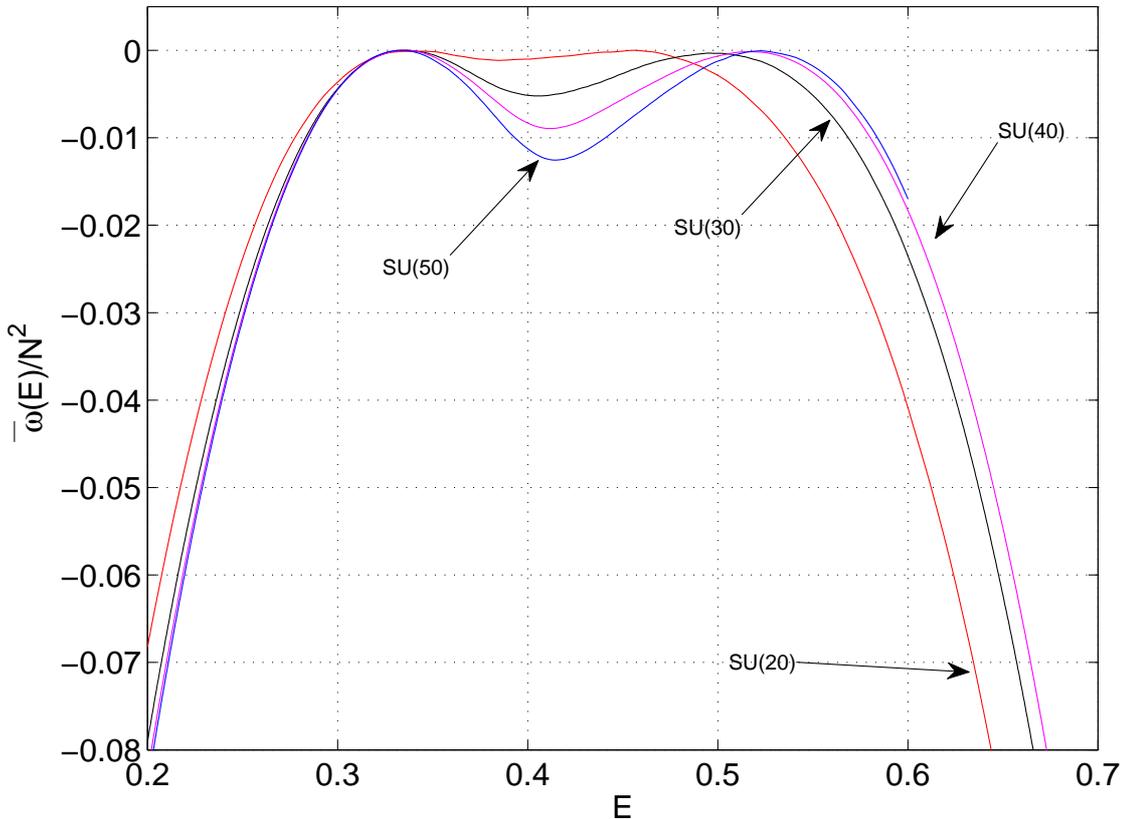}
\caption{Our most reliable estimates of
$\overline{\omega}(E)/N^2$. All results were obtained with 
$N_{\rm hit}=1$ and $\delta=0.005$, except for $N=50$ where
$\delta=0.0025$. The values of $\gamma$ were
$10^{-4},1.4\times10^{-5},2.5\times 10^{-5},3\times 10^{-5}$ for
$N=20,30,40,50$ respectively. For presentation purposes we shift the
maximum of $\overline{\omega}(E)$ to zero for each $N$.
Note that a smaller $E_{\rm max}$ was used for $N=50$.
\label{omega_final}
}
\end{figure}

\begin{figure}[tbh]
\includegraphics[width=15cm]{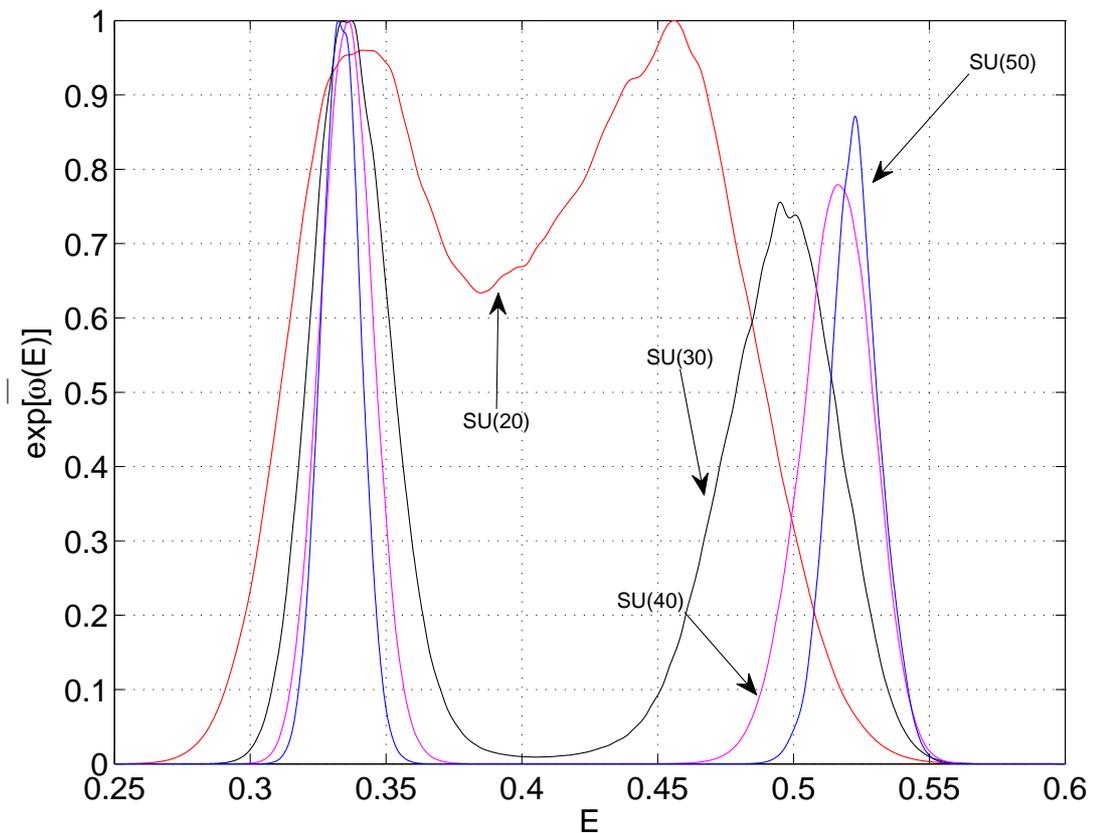}
\caption{Our most reliable estimates of $\exp(\overline{\omega}(E))$.
The parameters are as in Fig.~\protect\ref{omega_final}.
\label{P_final}
}
\end{figure}

To present our results
we first define the logarithm of the canonical
probability function, calculated at the transition coupling $b_t$:
\begin{equation}
\overline{\omega}(E) = \omega(E) + 12 \, N^2 \, b_t(N)
\,E. \label{omegabar}
\end{equation}
We write $b_t(N)$ in \Eq{omegabar} to emphasize that the transition
coupling $b_t$ depends on $N$.  Presenting $\overline{\omega}(E)$ and
not $\omega(E)$ makes the $N$ dependence more apparent. We calculate $\omega(E)$ by averaging over the measurements.
In Fig.~\ref{omega_final} we present 
$\overline{\omega}(E)/N^2$ for the gauge groups we
studied. 
The values of $b_t(N)$ used to generate this figure appear
in Table~\ref{tab_final_WL_res} and we discuss how we obtained them
below.  
In Fig.~\ref{P_final} we show the canonical probability
function itself, i.e. $\exp( \overline{\omega}(E))$. 
We do not show statistical errors in either figure since the meaning
of such an error for both $\overline{\omega}(E)$ or its exponent is
nontrivial: only differences of $\overline{\omega}(E)$ and ratios of
$\exp(\overline{\omega}(E))$ have physical meaning. 
Meaningful errors can be
computed using a simple scheme of error propagation but this is not
necessary for our purposes here.\footnote{%
For example, one can
estimate the statistical error in the ratio of
$\exp(\overline{\omega}(E))$ between adjacent values of $E$ and
propagate it in a stochastic manner to find the error in
$\exp(\overline{\omega}(E_1)-\overline{\omega}(E_2))$ for a finite
difference $|E_1-E_2|$.}
An indication of the size of the uncertainty is given,
however, from the ``wiggles'' in Fig.~\ref{P_final}.
Note that these are much larger than those in Fig.~\ref{omega_final},
because of the exponential enhancement.


The expected double-peak structure is clearly seen, yet the WL algorithm
has done its job by providing the density of states in the
intermediate regime. 
It is noteworthy that the dip in $\overline\omega/N^2$ grows with
increasing $N$. This is contrary to the usual behavior in field theories 
where the dip in this normalized quantity decreases as
the $N_{\rm dof}$ increases. 

\begin{figure}[htb]
\includegraphics[width=15cm]{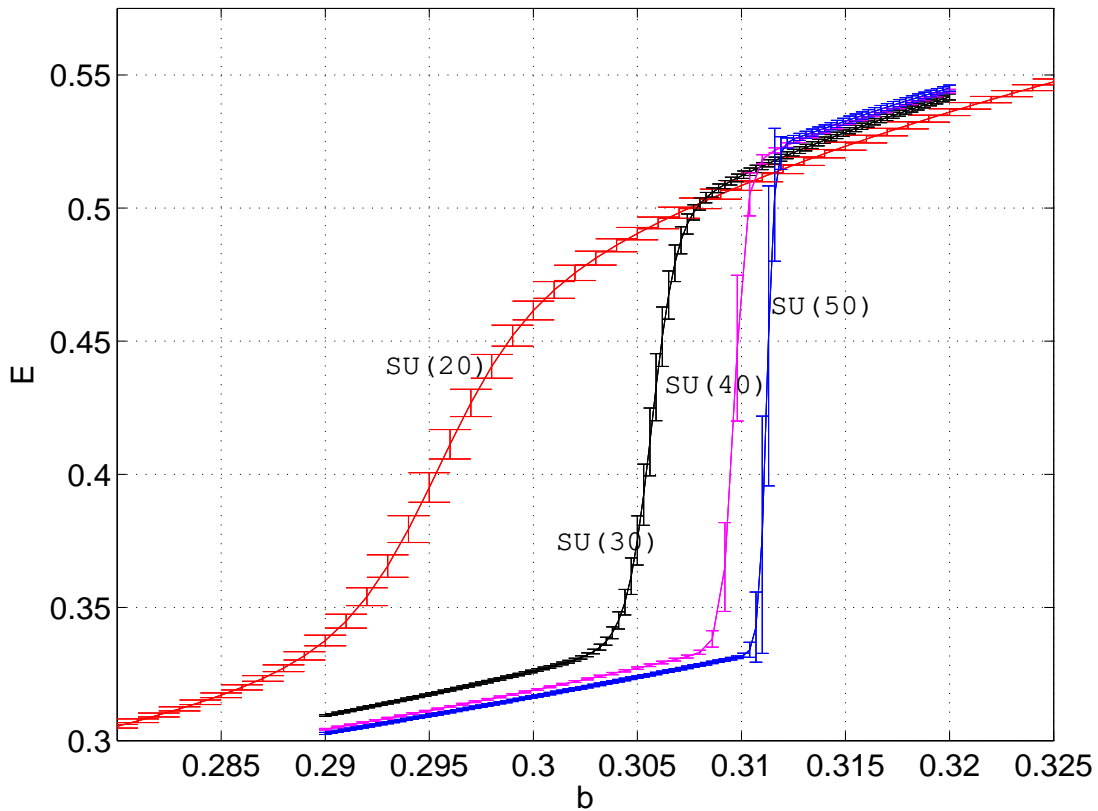}
\caption{Final results for ${\cal E}(b)$ obtained using WLR.
Error bars are shown at selected values of $b$, and are
highly correlated.
\label{E_WL}
}
\end{figure}

\begin{figure}[tbh]
\includegraphics[width=15cm]{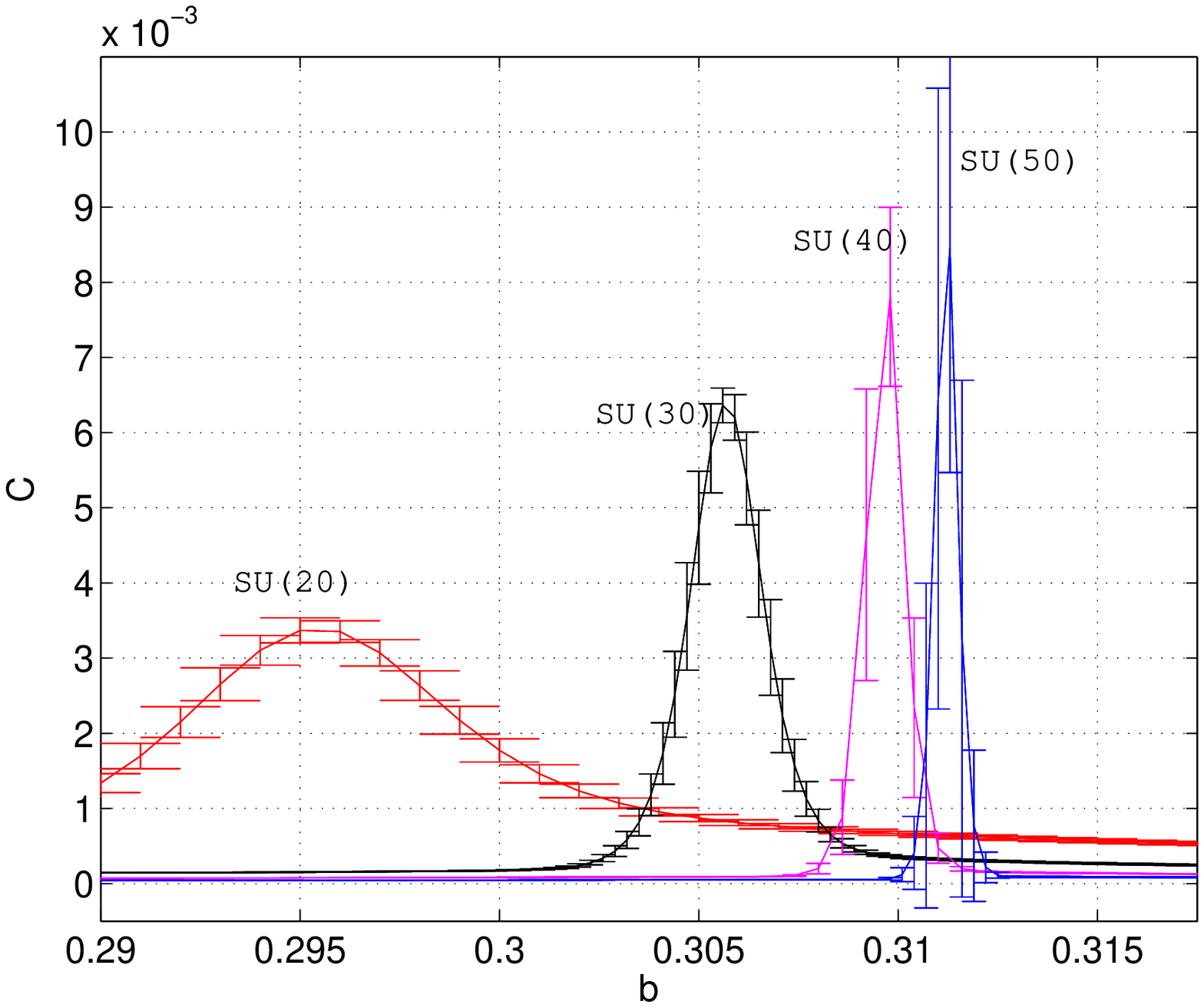}
\caption{Final results for ${\cal C}(b)$ obtained using WLR.
\label{C_WL}
}
\end{figure}

Using our estimates of $\omega(E)$ we can calculate 
the average action density ${\cal E}$
and the corresponding specific heat ${\cal C}$,
\begin{eqnarray}
{\cal E}(b) &=& Z^{-1}(b) \int dE \, \exp\left[ \omega(E) + 12N^2 b
E\right] E, \label{av_E}\\ 
{\cal C}(b) &=& Z^{-1}(b) \int dE \,
\exp\left[ \omega(E) + 12N^2 b E\right] \left(E-{\cal
E}(b)\right)^2\,, \\\label{av_C}
Z(b) &=& \int dE \,
\exp\left[ \omega(E) + 12N^2 b E\right] \,, \label{Z}
\end{eqnarray}
as a function of inverse `t Hooft coupling $b$, at least for
the range of $b$ where ${\cal E}$ lies well within our
range $[E_{\rm min},E_{\rm max}]$.
The results of doing so
are shown in Figs.~\ref{E_WL} and \ref{C_WL}.
All statistical errors are obtained
using the jackknife method, dropping single
  measurements of $\omega$ in turn from the average. 

We define the transition coupling $b_t$ to be the
location of the peak in ${\cal C}(b)$ and give the resulting values of
$b_t(N)$ in Tables~\ref{tab_res_su20}--\ref{tab_res_su50}.
The WL algorithm is clearly able to determine $b_t$ with high
accuracy (0.1-0.2\%). We note that error in $b_t$ is roughly
constant as $N$ increases, as long as we scale $\gamma$ so
that the tunneling rate stays approximately the same (as discussed
in the previous section). Since the number of full updates
is approximately the same for all $N$, this means that the
computational effort is growing proportional to $N^2$.
This is a much milder dependence than the exponential growth
required for canonical simulations.

All results for $b_t$ for a given $N$ are consistent, despite the
use of different values of the parameters of the algorithm.
As a further check we have used the Ferrenberg-Swendsen
multi-histogram reweighting method, which works well 
for $N=20,30$, but fails for $N\ge 40$.
The values of $b_t$ obtained with FSR are given
in Table~\ref{tab_FSR}, and agree within the very small errors
with those from the WL algorithm.

\begin{table}
\setlength{\tabcolsep}{4mm}
\begin{tabular}{cccccc}
\hline\hline $\gamma$ & $N_{\rm bin}$ & $N_{\rm meas}$ &
\parbox{3cm}{total number of tunnelings} & $b_t$ & ${\cal C}_t$\\
\hline 
$10^{-4}$ & $1000$ & $108$ & $\sim 140$ & $0.29585(45)$ &
$3.222(128)\times 10^{-3}$ \\ 
$10^{-4}$ & $1000^\star$ & $110$ & $\sim
400$ & $0.29597(21)$ & $3.336(55)\times 10^{-3}$ \\ 
$10^{-4}$ & $4000$ & $100$ & $\sim 120$ & $0.29544(37)$ &
 $3.399(55)\times 10^{-3}$ \\ 
$10^{-5}$ & $1000$ & $18$ & $\sim 80$ & $0.29509(37)$ &
$3.260(83)\times 10^{-3}$ \\ 
$10^{-6}$ & $1000$ & $20$ & $\sim 40$ &
$0.29585(13)$ & $3.402(35)\times 10^{-3}$ \\ \hline\hline
\end{tabular}
\caption{Results using WLR for $SU(20)$, using the range 
$E\in [0.1,0.7]$, and setting $\delta=0.005$ and $N_{\rm hit}=1$. 
The row denoted by a star was obtained by adding
explicit permutations to the update of the $SU(N)$ matrices, which
were found to be accepted $20\%$ of the time. For further details on
the importance of permutations we refer to Ref.~\cite{QEK_paper}.}
\label{tab_res_su20}
\end{table}

\begin{table}
\setlength{\tabcolsep}{3mm}
\begin{tabular}{ccccccc}
\hline\hline $\gamma$ & $N_{\rm hit}$ & $N_{\rm bin}$ & $N_{\rm meas}$
& \parbox{2.5cm}{total number of tunnelings} & $b_t$ & ${\cal C}_t$\\
\hline $2.25\times 10^{-4}$ & $1$ & $1500$ & $108$ & $\sim 140$ &
$0.30557(55)$ & $5.974(50)\times 10^{-3}$ \\ $2.25\times 10^{-4}$ &
$1$ & $50000$ & $100$ & $\sim 180$ & $0.30652(80)$ &
$5.051(260)\times 10^{-3}$ \\ $2.25\times 10^{-4}$ & $1740$ & $1500$ &
$114$ & $\sim 20$ & $0.30539(35)$ & $6.276(120)\times 10^{-3}$ \\
$1.4\times 10^{-5}$ & $1$ & $1500$ & $102$ & $\sim 60$ &
$0.30569(17)$ & $6.395(230)\times 10^{-3}$ \\ $1.4\times 10^{-4}$ &
$1$ & $1500$ & $110$ & $\sim 20$ & $0.30551(30)$ & $6.453(120)\times
10^{-3}$ \\ \hline\hline
\end{tabular}
\caption{Results from WLR for $SU(30)$, using the range $E\in [0.1,0.7]$.
All calculations were done with $\delta =0.005$ except for that
presented in the last row, for which $\delta=0.0008$.  }
\label{tab_res_su30}
\end{table}

\begin{table}
\setlength{\tabcolsep}{4mm}
\begin{tabular}{ccccc}
\hline\hline $\gamma$ & $N_{\rm meas}$ & \parbox{3cm}{total number of
tunnelings} & $b_t$ & ${\cal C}_t$\\ \hline $4\times 10^{-4}$ & $151$
& $\sim 850$ & $0.30965(75)$ & $6.67(30)\times 10^{-3}$ \\
$2.5\times 10^{-5}$ & $355$ & $\sim 75$ & $0.30968(20)$ &
$8.24(10)\times 10^{-3}$ \\ \hline\hline
\end{tabular}
\caption{Results from WLR for $SU(40)$, using the range $E\in
[0.1,0.7]$, and obtained with $\delta=0.005$, $N_{\rm bin}=1875$, 
and $N_{\rm hit}=1$.  }
\label{tab_res_su40}
\end{table}

\begin{table}
\setlength{\tabcolsep}{4mm}
\begin{tabular}{ccccc}
\hline\hline $\gamma$ & $N_{\rm meas}$ & \parbox{3cm}{total number of
tunnelings} & $b_t$ & ${\cal C}_t$\\ \hline $3\times 10^{-5}$ & $108$
& $\sim 45$ & $0.31119(19)$ & $9.02(12)\times 10^{-3}$ \\
\hline\hline
\end{tabular}
\caption{Results from the WLR for $SU(50)$, using the range 
$E\in [0.1,0.6]$, and obtained with $\delta=0.0025$, $N_{\rm bin}=2100$, 
and $N_{\rm hit}=1$.  }
\label{tab_res_su50}
\end{table}

\begin{table}
\setlength{\tabcolsep}{4mm}
\begin{tabular}{cccc}
\hline\hline $N$ & Reweighting using & $b_t$ & ${\cal C}_t$\\ \hline
$20$ & $40$ histograms & $0.295980(48)$ & $3.32(3)\times 10^{-3}$ \\
$30$ & $17$ histograms & $0.30551(30)$ & $6.45(5)\times 10^{-3}$\\
\hline\hline
\end{tabular}
\caption{Results from FSR multi-histogram reweighting. For $N=20(30)$
each histogram contains an average of $10^4$ ($5\times 10^4$) measurements,
separated by 5 full model updates from each other.  }
\label{tab_FSR}
\end{table}

\begin{figure}[htb]
\includegraphics[width=15cm]{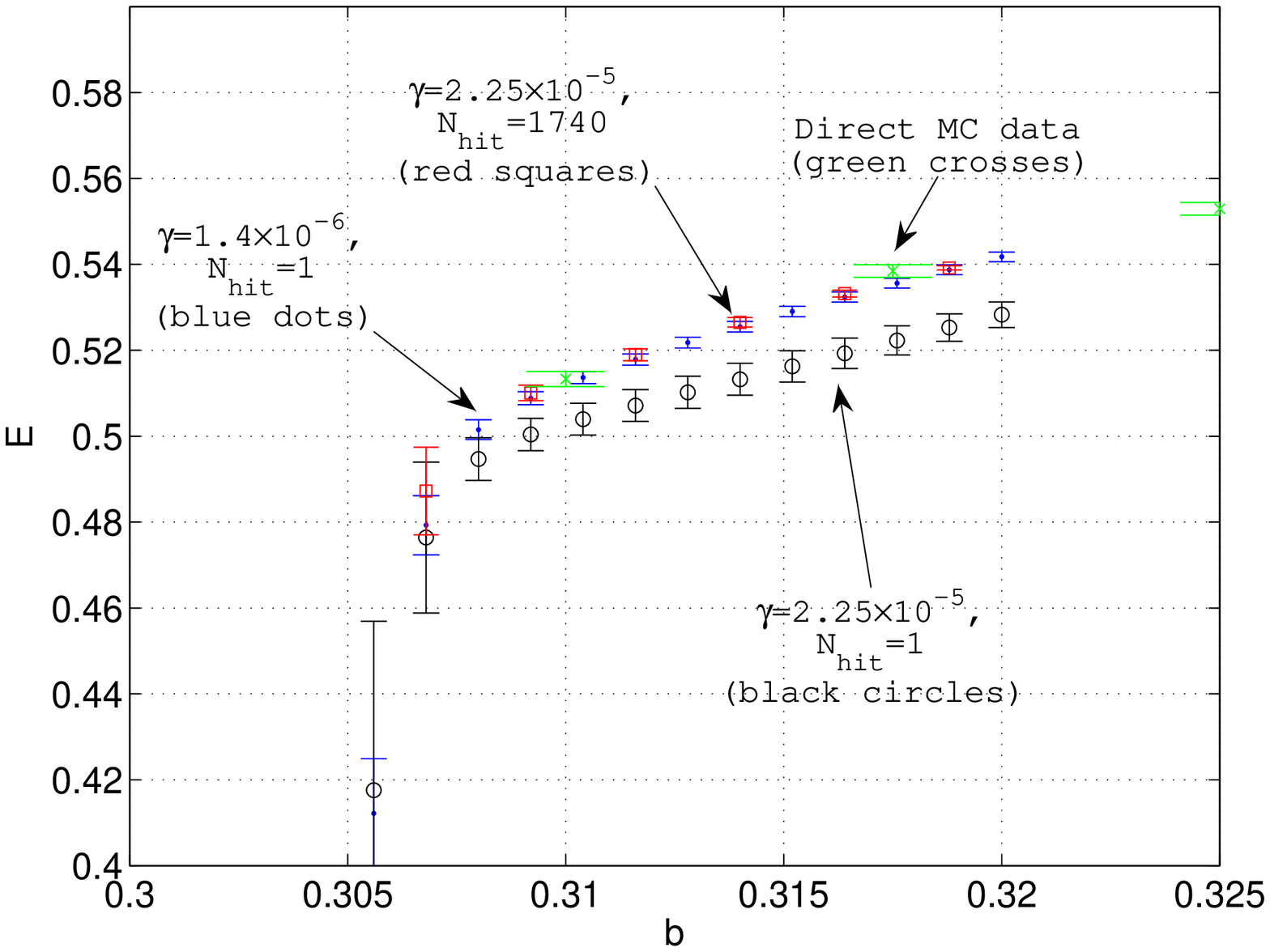}
\caption{Comparison of WLR results for the average action density for $SU(30)$,
with direct measurements from standard MC simulations. The systematic error discussed in Section~\ref{Nhitsmall} is clearly seen.
\label{sys_E}
}
\end{figure}

The situation is different for the results for 
the peak value of the specific heat, ${\cal C}_t\equiv {\cal C}(b=b_t)$. 
Although results in Tab.~\ref{tab_res_su20} for $N=20$ are consistent,
those in Tabs.~\ref{tab_res_su30} and \ref{tab_res_su40} for $N=30$
and $40$ are not.
In addition, we find discrepancies with results from canonical MC
simulations. These are exemplified by 
Fig.~\ref{sys_E}, 
where we present the estimates of 
${\cal E}(b\stackrel{>}{_\sim}0.305)$ from WLR for $SU(30)$ together with the
values obtained from direct MC simulations.
What we find is that we obtain agreement with FSR and/or
canonical MC results only if either $\gamma$ is small enough
or $N_{\rm hit}$ is large. This is presumably the realization of the systematic error discussed in
Section~\ref{sec_tune} and  illustrates the importance of having results at more than
one value of $\gamma$ or $N_{\rm hit}$.

\section{Summary}
\label{sec_summary}

In this paper we present an implementation of 
a variant of the Wang-Landau
reweighting algorithm in the context of $SU(N)$ lattice gauge
theories.\footnote{%
The original WL algorithm was implimented for $U(1)$ 
gauge theory in Ref.~\cite{BB}, and used to provide an input
weighting function for a multicanonical simulation.}

This algorithm was introduced in the field of statistical
mechanics to calculate the density of states of discrete spin
systems. We use a generalization of the original algorithm to systems
with continuous degrees of freedom and apply it to a matrix model that
is obtained by quenched reduction from four-dimensional $SU(N)$
lattice gauge theory. This matrix model consists of four $SU(N)$
matrices with interactions governed by the `t Hooft coupling
$\lambda$, and has a first-order strong-to-weak coupling
phase transition in its large-$N$
limit at $\lambda=\lambda_t$. An accurate measurement of $\lambda_t$
at $N=\infty$ is what we aimed to achieve using WLR.

\begin{figure}[htb]
\centerline{ \includegraphics[width=15cm]{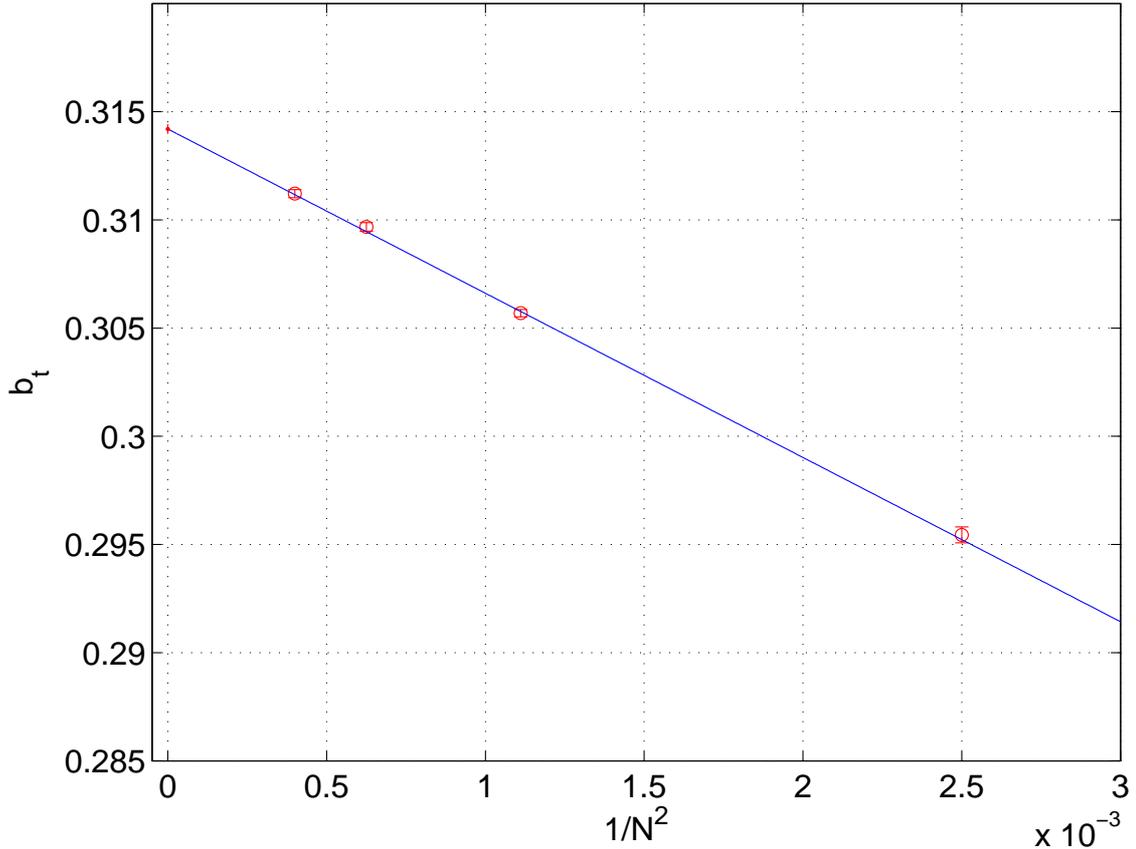} }
\caption{The strong-to-weak transition coupling, $b_t=1/\lambda_t$,
plotted versus $1/N^2$. [Red] squares show our results using the
Wang-Landau algorithm from Table~\ref{tab_final_WL_res}.  The solid
blue curve is the linear fit described in the text (with parameters
listed in the first row of
Table~\ref{tab_res_fit}).  }
\label{final_WL_res}
\end{figure}

Our variant of the WL algorithm does not extrapolate
the flucturations in the Boltzman weights towards zero, but rather
retains these fluctuations at a small, non-zero value, in order to
maintain tunneling at a first-order transition.
Assuming these fluctuations are symmetric around zero, we can systematically
estimate the error in the density of states and in derived quantities
such as the specific heat.
We have studied the systematic errors associated with chosing the various
parameters of the algorithm.
Our most reliable WL estimates
of $\lambda_t$ for gauge groups with $N=20,30,40,50$ are summarized in
Table~\ref{tab_final_WL_res} and plotted in Fig.~\ref{final_WL_res}
versus $1/N^2$.
\begin{table}
\setlength{\tabcolsep}{4mm}
\begin{tabular}{ccccc}
\hline\hline $N$ & $20$ & $30$ & $40$ & $50$ \\ \hline
$b_t(N)=(\lambda_t)^{-1}_N$ & $0.29544(37)$ & $0.30569(17)$ &
$0.30968(20)$ & $0.31121(19)$ \\ \hline\hline
\end{tabular}
\caption{A summary of our most reliable WL results for
$b_t(N)=(\lambda_t)^{-1}_N$. \label{tab_final_WL_res} }
\end{table}
We fit our data to the form
\begin{equation}
(\lambda_t)^{-1}_N = (\lambda_t)^{-1}_\infty + \frac{A}{N} +
\frac{B}{N^2}, \label{fit}
\end{equation}
and present the results of these fits in Table~\ref{tab_res_fit}. We also plot the
result of the linear fit in $1/N^2$ (i.e. the fit with $A=0$ whose results are presented in the first row of
Table~\ref{tab_res_fit}) in Fig.~\ref{final_WL_res}.
\begin{table}
\setlength{\tabcolsep}{4mm}
\begin{tabular}{ccccc}
\hline\hline Type of fit & $(\lambda_t)^{-1}_\infty$ & $A$ & $B$ &
$\chi^2/{\rm d.o.f.}$\\ \hline $A=0,B\neq 0$ & 0.3142(2) & -- &
-7.59(18) & $1.45/2$ \\ $A\neq 0,B\neq 0$ & 0.3148(10) & -0.037(65) &
-7.06(97) & 1.1/1 \\ \hline\hline
\end{tabular}
\caption{The fit parameters $(\lambda_t)^{-1}_\infty$, $A$, and $B$,
obtained from fitting the Wang-Landau data in
Table~\ref{tab_final_WL_res} to the form \Eq{fit}. The first row shows results from a fit linear in $1/N^2$ (i.e. with $A=0$). \label{tab_res_fit}}
\end{table}

We find that in the large-$N$ limit $(\lambda_t)^{-1}_\infty =
 0.3142(2)-0.3148(10)$, depending on the way we fit. These results are
 many standard deviations away from the value of $(\lambda)^{-1}_{\rm
 Bulk}\simeq 0.36$ where the strongly first order `bulk' transition
 takes place in four-dimensional $SU(\infty)$ lattice gauge
 theories. This discrepancy is one of several pieces of evidence
 adduced in Ref.~\cite{QEK_paper} for the
 breakdown of large-$N$ quenched reduction in four-dimensional $SU(N)$ lattice gauge
 theories, and we refer the reader to that paper for further
 discussion. Such a discrepancy was not seen in past explorations of
 the matrix model partly because the phase transition is so strong
 that it is very hard to measure its transition coupling by
 conventional means. The Wang-Landau algorithm allowed us to solve this
 problem and to determine that there is a discrepancy.
 We conclude that the Wang-Landau algorithm can be a useful and
 feasible way to study $SU(N)$ lattice gauge theories.

\section*{Acknowledgments}
We acknowledge A.~Bazavov, B.~Berg and S.~Trebst for useful discussions.
This work was supported in part by the U.S. Department of
Energy under Grant No. DE-FG02-96ER40956.

\appendix

\section{Algorithms for simulating 
quenched, reduced $SU(N)$ lattice gauge theory}
\label{app_alg}

In our standard (non Wang-Landau) MC simulations of the model we used several different
algorithms to update the matrices $V_\mu$. 
We use a standard Metropolis algorithm (M),
a ``hybrid'' heat-bath $+$ Metropolis (HM), 
and a full heat-bath (HB)---the latter including
different types of over-relaxations. For the M and
HM algorithms we generate a set of random $SU(2)$ matrices at the
beginning of the run and keep them in memory. For each matrix in this
list we add to the list its inverse. The M algorithm is completely
standard. We randomly choose an $SU(2)$ matrix $u$ 
from the list, extend it to an $SU(N)$
matrix by adding $1$'s along the diagonal,
update $V_\mu \to u \, V_\mu$ 
and accept this proposed update with the usual Metropolis probability.
This is repeated five times for equilibration.
This process is then repeated, following Cabibbo and Marinari~\cite{CM}, 
for each of the [$N(N-1)/2$] $SU(2)$ subgroups of $SU(N)$ in turn, 
and for each $V_\mu$ in turn.

We now describe the other algorithms, which are less standard.

\subsection{Hybrid Heat-bath Algorithm}
\label{HM}
Here we use the prescription suggested in Ref.~\cite{FH} to make
the action linear in the link matrices $U_\mu$.
This requires a Hubbard-Stratonovich Gaussian field $Q_{\mu\nu}$ for each
plaquette. It results in
an effective action $A_{\rm eff}(U_\mu,Q_{\mu\nu})$
that is quadratic in $Q_{\mu\nu}$ and linear in $U_\mu$---and thus
quadratic in $V_\mu$. 
We then update one of the $V_\mu$ as follows:
\begin{enumerate}
\item First update the matrices $Q_{\mu,\nu}$. This update is trivial
since $Q$ has a (shifted) Gaussian distribution.
\item Update $V_\mu$ as in the M algorithm using all $SU(2)$
subgroups but now with the action $A_{\rm eff}$. 
\end{enumerate}
This is repeated in turn for each of the links.

\subsection{Heatbath and over-relaxation algorithms}
\label{HB}

The heatbath algorithm requires the use of two auxiliary fields
in order to obtain an action which is linear in the $V_\mu$. 
It is not quite as simple as applying the approach of Ref.~\cite{FH}
twice, and so we give some details. 

We begin by recalling the action 
\begin{equation}
A = 2 N \sum_{\mu<\nu} \Re\Tr\left(U_\mu U_\nu U_\mu^\dagger
U_\nu^\dagger\right) \,,
\end{equation}
where $U_\mu = V_\mu \Lambda_\mu V_\mu^\dagger$
and $(\Lambda_\mu)_{ab} = \delta_{ab} \exp(i p^a_{\mu})$.
We next define two sets of unitary matrices
\begin{equation}
A_{\mu\nu} \equiv V_\mu^\dagger V_\nu =A_{\nu\mu}^\dagger\,, 
\qquad 
B_{\mu\nu} \equiv A_{\mu\nu} \Lambda_\nu A_{\mu\nu}^\dagger
\ne B_{\nu\mu}^\dagger\,,
\qquad (\mu\ne\nu)\,,
\label{eq:ABdef}
\end{equation}
in terms of which the action can be written as
\begin{eqnarray}
A &=& 2 N \sum_{\mu<\nu} \Re\Tr\left(A_{\mu\nu} \Lambda_\nu
           A_{\mu\nu}^\dagger \Lambda_\mu^\dagger A_{\mu\nu}
           \Lambda_\nu^\dagger A_{\mu\nu}^\dagger \Lambda_\mu\right)
           \,,\label{eq:SfromA} 
\\ 
&=& 2 N \sum_{\mu<\nu}
           \Re\Tr\left(B_{\mu\nu} \Lambda_\mu^\dagger
           B_{\mu\nu}^\dagger \Lambda_\mu\right) \,.\label{eq:SfromB}
\end{eqnarray}
For each plaquette, i.e. for each $\mu<\nu$ we introduce auxiliary complex
fields $\tQ_{\mu\nu}$ and $\tP_{\mu\nu}$, with Boltzmann weights 
\begin{equation}
\exp[- b{N} \, \Tr(\tQ_{\mu\nu}\tQ_{\mu\nu}^\dagger)
     - b N \, \Tr(\tP_{\mu\nu}\tP_{\mu\nu}^\dagger)] \,.
\end{equation}
These are then shifted as follows:
\begin{eqnarray}
Q_{\mu\nu} &=& \tQ_{\mu\nu} +
\left\{B_{\mu\nu},\Lambda_\mu^\dagger\right\} \,,\label{eq:Qdef} \\
Q^\Lambda_{\mu\nu} &=& \left\{Q_{\mu\nu},\Lambda_\mu\right\} =
\left\{\tQ_{\mu\nu},\Lambda_\mu\right\} + 2 B_{\mu\nu} + \Lambda_\mu
B_{\mu\nu} \Lambda_\mu^\dagger + \Lambda_\mu^\dagger B_{\mu\nu}
\Lambda_\mu \,,\label{eq:QDdef} \\ P_{\mu\nu} &=& \tP_{\mu\nu} +
A_{\mu\nu}\Lambda_\nu^\dagger + {Q^\Lambda_{\mu\nu}}^\dagger
A_{\mu\nu} \,.\label{eq:Pdef}
\end{eqnarray}
The staple-like quantity $X_{\mu\nu}$ can then be calculated:
\begin{equation}
X_{\mu\nu} = \Lambda_\nu^\dagger P_{\mu\nu}^\dagger +
P_{\mu\nu}^\dagger {Q^\Lambda_{\mu\nu}}^\dagger \,,\qquad 
X_{\nu\mu} = X_{\mu\nu}^\dagger \,,\qquad
(\mu < \nu)
\label{eq:Xdef}
\end{equation}
Finally, the action can be written in a form suitable for a heatbath or
overrelaxed update:
\begin{eqnarray}
A' &=& A - N \sum_{\mu<\nu} \left[
                   \Tr(\tQ_{\mu\nu}\tQ_{\mu\nu}^\dagger) +
                   \Tr(\tP_{\mu\nu}\tP_{\mu\nu}^\dagger)
                   +\textrm{const.}\right] 
\\
   &=&     N  \sum_{\mu\ne\nu} \Tr\left(V_\mu X_{\nu\mu}
                   V_\nu^\dagger\right) 
\nonumber \\ 
   & &    - N \sum_{\mu<\nu}
                   \left[\Tr(Q_{\mu\nu} Q_{\mu\nu}^\dagger) +
                    \Tr(Q_{\mu\nu}^\Lambda
                   {Q_{\mu\nu}^{\Lambda}}^\dagger)
                   + \Tr(P_{\mu\nu} P_{\mu\nu}^\dagger) \right] 
\label{eq:Sprime2} \,.
\end{eqnarray}

In summary, to evaluate an observable that depends only on the gauge
fields one can use
\begin{eqnarray}
\langle O(U)\rangle &=& \frac1{Z} \int \prod_\mu DV_\mu \, e^{b \, A}
\, O(U) 
\\ 
&=& \frac1{Z'} \int \prod_\mu DV_\mu \prod_{\mu<\nu}
D\tQ_{\mu\nu} D\tQ_{\mu\nu}^\dagger D\tP_{\mu\nu}
D\tP_{\mu\nu}^\dagger\, e^{b\, A'}\, O(U) 
\\ 
&=& \frac1{Z'} \int
\prod_\mu DV_\mu \prod_{\mu<\nu} DQ_{\mu\nu} DQ_{\mu\nu}^\dagger
DP_{\mu\nu} DP_{\mu\nu}^\dagger\, e^{b\, A'}\, O(U)\,,
\label{eq:Zprime2}
\end{eqnarray}
where
\begin{equation}
Z' = \int \prod_\mu DV_\mu \prod_{\mu<\nu} D\tQ_{\mu\nu}
D\tQ_{\mu\nu}^\dagger D\tP_{\mu\nu} D\tP_{\mu\nu}^\dagger\, e^{b \, A'}
\,. \label{Zprime3}
\end{equation}
The form (\ref{eq:Zprime2}), together with the expression for $A'$ in
eq.~(\ref{eq:Sprime2}), shows how the auxiliary fields decouple the
$V$'s.  To get the correctly distributed $V$'s and $P$'s, one
can update $V_\mu$ using the ``staple'' part of the action: 
\begin{equation}
A_{\rm staple}(V_\mu) = 2 N \sum_{\nu\ne\mu} \Re\Tr(V_\mu X_{\nu\mu}
V_\nu^\dagger)\,,
\end{equation}
where we stress that in this case the sum is now only over $\nu$.
This form is suitable for a heat-bath update, which we implement
in each $SU(2)$ subgroup in turn.

To generate the correct distribution of the $P$'s is
straightforward. Given the $Q$'s and $V$'s, one can generate $\tP$'s
with the Gaussian measure and make the shifts given in
eq.~(\ref{eq:Pdef}). This leads to the correct linear and quadratic
terms in $P_{\mu\nu}$ in the action of eq.~(\ref{eq:Sprime2}).

To update the $Q$'s one must be more careful. Simply generating
$\tQ$'s with Gaussian measure and using the shift of
eq.~(\ref{eq:Qdef}) leads to the {\em wrong} distribution: neither the
quadratic or the linear terms in eq.~(\ref{eq:Sprime2}) are
reproduced. Instead, one should ``complete the square'' using the
terms that are present in eq.~(\ref{eq:Sprime2}).  To do so requires
that one first generate the $\tQ$'s using the Gaussian measure, 
but then shifts and rescales as follows:
\begin{eqnarray}
(Q_{\mu\nu})_{ab} &=& \alpha_{ab} (\tQ_{\mu\nu})_{ab} + \alpha_{ab}^2
(\{A_{\mu\nu} P_{\mu\nu}^\dagger,\Lambda_\mu^\dagger\})_{ab}\,, \\
\alpha_{ab} &=& \frac1{\sqrt{3 + 2 \cos(p_{\mu}^a-p_\mu^b)}}\,.
\end{eqnarray}

Thus the structure of the algorithm is as follows.
One begins with an initial choice of $V$'s {\em and} $Q$'s.
Then one can update all the $P$'s, update all the $Q$'s, 
and finally update the $V$'s 
(updating all directions for given $Q$'s and $P$'s). To return
to the beginning of the loop one needs to store not
only the $V$'s but also the $Q$'s.
One could also interchange the ordering and roles of the
$P$'s and $Q$'s.

Once one has an effective action that is linear in the $U(N)$ matrices
$V_\mu$ the way is open for over-relaxation algorithms. We have thus
implemented both an over-relaxation in all the $SU(2)$ subgroups of
$SU(N)$ as well as a full $SU(N)$ over-relaxation of the type
described in \cite{OR}.

\subsection{Update scheme in the Wang-Landau algorithm}
\label{alg_WL}

In the WL algorithm we need to propose changes to the $V_\mu$.
This we do as in the Metropolis and HM algorithms,
i.e. one $SU(2)$ subgroup at a time. Such an update is
what we refer to as a ``hit'', so if
$N_{\rm hit}=1$ we change 
$\omega_t(E)$ after each individual $SU(2)$ multiplication.
We also checked that updating $\omega(E)$ in between full $SU(N)$ updates
(for all four $V_\mu$) 
gives similar results---this gives $N_{\rm hit}=2 \times N(N-1)$.
In either case,
we call a ``full update'' the update of all $SU(2)$ subgroups for all
four links.

Finally, we have considered an extra type of update
which permutes the angles $p^a_\mu \leftrightarrow p^b_\mu$ for
randomly chosen pairs of indices $a$ and $b$. This was motivated by
the importance of permutations in this quenched-reduced model~\cite{QEK_paper}.

\section{More practical issues}
\label{app_more}

In this section we list several practical issues relevant to
the implementation of WLR.

\subsection{A initial guess for $\omega_{t=0}(E)$}
\label{app_guess}

We suggest performing the first implementation of the WLR with a
relatively low value of $N_{\rm dof}=N^{(0)}_{\rm dof}$. This run can
begin with a `blind' initial guess of $\omega_0(E;N^{(0)}_{\rm
dof})=0$. We then found it useful to appropriately scale the best
estimate of $\omega(E,N^{(0)}_{\rm dof})$ with $N_{\rm dof}$, so as to
use it as a good initial guess for $N_{\rm dof}>N^{(0)}_{\rm
dof}$. Since $\omega(E)$ is an extensive quantity this means
setting
\begin{equation}
\omega_{t=0}(E;N^{(1)}_{\rm dof}) = \frac{N^{(1)}_{\rm
dof}}{N^{(2)}_{\rm dof}} \, \omega_{t=\infty}(E;N^{(2)}_{\rm dof}).
\end{equation}
In our case, with $N_{\rm dof}\sim N^2$, this corresponds to
$\omega_{t=0}(E;N^2_1) = \left(\frac{N_1}{N_0}\right)^2 \,
\omega_{t=\infty}(E;N^2_0)$. The generalization for a field theory in
a finite lattice volume is obvious. We found that this procedure
shortens the initial stage of WLR considerably.

\subsection{Boundary effects}
\label{app_boundary}

As we mention in Section~\ref{sec_tune} it is useful to use WLR in a
subset of the full range of $E$.
This is sufficient if at the values of $b$ of interest, the
average action density ${\cal E}$ is localized far from the regime's
boundaries. 

This modification complicates the theoretical analysis
of Sec.~\ref{sec_WLA_theory} in a way
we only partially addressed.

The presence of the boundaries raises two practical questions
\begin{enumerate}
\item
What do we do when the update $E_{\rm old}\to E_{\rm new}$ results in a
value $E_{\rm new}$ which is outside of the region $[E_{\rm
min},E_{\rm max}]$ ?  
\item
What do we do when we update $\omega_t\to\omega_{t+1}$ and the
update function $F_{\delta}(E,E_t)$ extends outside the desired region?
\end{enumerate}
In this work we generalize the proposals of Ref.~\cite{bound_paps}.
The answer to the first question is that we reject $E_{\rm new}$
and thus perforce stay inside the desired region.  This means setting $E_{\rm new}=E_{\rm old}$ and updating $\omega_t(E_{\rm new})$ as in a regular update. 
This is the standard approach, which one can understand as follows.
If one were simulating the full range of $E$
then every time one left the range $[E_{\rm min},E_{\rm max}]$ one would eventually return. We are just dropping the MC-time history of the WL algorithm for which $E$ was outside the desired range. Since at time $t$ the WL algorithm updates $\omega_t(E)$ only around $E_t$, performing WLR in this way gives an estimate for $\omega(E)$ which is correct away from the boundaries $E=E_{\rm min,max}$.

Our answer to the second question is to ``reflect'' the part of
$F_\delta(E, E_t)$ which lies outside of the desired range back
into the range. 
The precise definition of the reflection is as follows.
For all $E$ (even those outside $[E_{\rm min},E_{\rm max}]$) perform
\begin{equation}
\omega(E') \to \omega(E') + \gamma F(E,E_{\rm old})\,,
\label{reflc}
\end{equation}
with
\begin{equation}
E' = \left\{
\begin{array}{lc}
E & \quad \qquad {\rm for} \quad E \in [E_{\rm min},E_{\rm max}],\\
2E_{\rm min}-E & {\rm for} \quad E<E_{\rm min},\\ 2E_{\rm max}-E &
{\rm for} \quad E>E_{\rm max}.
\end{array}
\right. \label{bound}
\end{equation}
Here we assume that $E'$ always obeys $E'\in [E_{\rm min},E_{\rm
max}]$. This is valid as long as the interval size
$(E_{\rm max}-E_{\rm min})$ is larger than the average change $|E_{\rm
new}-E_{\rm old}|$, which is very well satisfied in practice.\footnote{ 
We note in passing that once $F_\delta$ drops below
a certain size ($10^{-8}$ was our choice) we set it to zero. This saves calculation time since one needs to update $\omega(E)$ only for a subset of $E$, and also avoids the problem of
double and higher-order reflections.}

The reflected update has two important properties.
First, it maintains the result that the area $\int dE\, F_\delta(E,E_t)$ 
is independent of $E_t$. Second, for our Gaussian choice
of $F_\delta$, the definition remains symmetric:
$F_\delta(E,E_t)=F_\delta(E_t,E)$.

We find that if one does not perform the updates that corresponds to
last two rows in \Eq{bound}, then the WLR fails to converge, and
effectively overestimates $\omega(E)$ near the boundaries. This is
easy to understand: if we do not reflect the contribution of the
Gaussian in \Eq{reflc} then effectively the update to $\omega(E)$
close to the boundary is smaller than it would be in the absence of
the boundary.\footnote{To demonstrate this one can take the limit of
$\delta\to 0$ and reproduce the case discussed in \cite{bound_paps}.
We also note that maintaining the area under the $F_\delta$
plays an important role in the theoretical analysis of
Sec.~\ref{sec_WLA_theory}.}
Thus the force that drives one away from the boundary is too weak and
one spends too much time updating $\omega_t(E)$ near the boundary.

\subsection{Storing the functions $\omega_t(E)$ and $h_t(E)$}
\label{app_store}

Despite the continuous fashion in which we implement the WL
algorithm, one
still needs to store the functions $\omega_t(E)$ and $h_t(E)$ in
memory, as well as to update them. We do this by dividing the range
$[E_{\rm min},E_{\rm max}]$ into $N_{\rm bin}$ bins. The criterion
that determines the bin size 
$\delta E = (E_{\rm max}-E_{\rm min})/N_{\rm bin}$ is 
the same as that for $\delta$ (see Sec.~\ref{tuning_delta}).
In order for the error in the numerical evaluation of the
integral \Eq{OE} by binning remain small, one must have
\begin{equation}
\delta E \ll \sigma,,
\end{equation}
where $\sigma$ is the width in $E$ of the canonical distribution function
$P_C(E)$ of \Eq{p_c_def}.
Assuming that $\bar\omega(E)/N^2$ is quadratic about the peak
and has a good $N\to\infty$ limit (which appears to hold for
the ``outside'' branches of the peaks---see Fig.~\ref{omega_final}),
then one can show that $\sigma\propto 1/N$.
Thus one must increase the number of bins as $N_{\rm bin}\propto N$,
as we have done (see Tables~\ref{tab_res_su20}-\ref{tab_res_su50}).

Another criterion one might consider using is to
enforce a relationship between $\delta E$ and the average step size.
Naively one might think that the bins should be smaller than
the average step size, so that discretization effects do not
hinder the motion in $E$-space. 
This would be an onerous requirement, since our
step size, which is $\sim (1-5)\times 10^{-4}$, would require
significantly more bins than we use in most simulations.
In fact, it turns out that the acceptance rate, and
thus the motion through $E$-space is almost
independent of the bin size. We have seen this numerically
for $SU(30)$, where we have one simulation with 50000 bins.
But one can also understand this analytically if
the step size is much smaller than $\sigma$, which is the
case for our simulations. We do not present the derivation,
but the essential point is that with bins much larger than
the step size the smaller acceptance when jumping
between bins (when the step is to lower $E$) is
exactly counterbalanced by the free motion
(without rejection) within the bins.

\end{document}